\documentclass{article}

\usepackage{arxiv}

\usepackage{braket}
\usepackage{anyfontsize}
\usepackage{subcaption}
\usepackage[bottom]{footmisc}
\usepackage{arxiv}
\usepackage{amsmath}
\usepackage[utf8]{inputenc} 
\usepackage[T1]{fontenc}    
\usepackage{hyperref}       
\usepackage{url}            
\usepackage{booktabs}       
\usepackage{amsfonts}       
\usepackage{nicefrac}       
\usepackage{microtype}      
\usepackage{lipsum}		
\usepackage{graphicx}
\usepackage{doi}

\usepackage{braket}

\title{Quantum Data Centres: \\
Why Entanglement Changes Everything
\thanks{This version is licensed under CC BY.}}

\author{ A. S. Cacciapuoti, C. Pellitteri and J. Illiano, L. d'Avossa and F. Mazza, S. Chen, M. Caleffi\\The authors are with the \href{www.quantuminternet.it}{www.QuantumInternet.it}  Group, University of Naples Federico II, Naples, 80125 Italy\\ \textbf{Author for correspondence:}Angela Sara Cacciapuoti\\
{angelasara.cacciapuoti@unina.it}}
\date{}
\renewcommand{\shorttitle}{\textit{arXiv} Template}

\hypersetup{
pdftitle={A template for the arxiv style},
pdfsubject={q-bio.NC, q-bio.QM},
pdfauthor={David S.~Hippocampus, Elias D.~Striatum},
pdfkeywords={First keyword, Second keyword, More},
}

\begin{document}
\maketitle

\begin{abstract}
The Quantum Internet is key for distributed quantum computing, by interconnecting multiple quantum processors into a virtual quantum computation system. This allows to scale the number of qubits, by overcoming the inherent limitations of noisy-intermediate-scale quantum (NISQ) devices. Thus, the Quantum Internet is the foundation for large-scale, fault-tolerant quantum computation. Among the distributed architectures, Quantum Data Centres emerge as the most viable in the medium-term, since they integrate multiple quantum processors within a localized network infrastructure, by allowing modular design of quantum networking.
We analyze the physical and topological constraints of Quantum Data Centres, by emphasizing the role of entanglement orchestrators in dynamically reconfiguring network topologies through local operations. We examine the major hardware challenge of quantum transduction, essential for interfacing heterogeneous quantum systems. Furthermore, we explore how interconnecting multiple Quantum Data Centres could enable large-scale quantum networks. We discuss the topological constraints of such a scaling and identify open challenges, including entanglement routing and synchronization. The carried analysis positions Quantum Data Centres as both a practical implementation platform and strategic framework for the future Quantum Internet.
\end{abstract}


\section{Introduction}
\label{sec:01}
 Quantum computing is one of the most promising applications of quantum technologies, gathering significant investments from leading technology firms such as Google, IBM, and Amazon. Indeed, the promises of quantum computing span across all major industrial sectors, with particular emphasis on sustainable energy optimization, an area where quantum-enabled advances are expected to drive significant progress in the coming years. Furthermore, as Shor demonstrated in 1994 \cite{Shor1994}, quantum computing has the potential to break Rivest–Shamir–Adleman (RSA), the most widely cryptographic protocol used on Internet nowadays. And for its implementation, thousands of qubits are required \cite{CacCalTaf-20, CalAmoFer-22}. More into details, breaking a $2048$-bit RSA key requires roughly $4\text{,}099$ \textit{logical} (or fault-tolerant) \cite{Bea-03} qubits. The mapping between \textit{logical} qubits and \textit{physical qubits} depends on the specific adopted technology (or qubit implementation) and usually implies a huge overhead. By considering the different qubit implementations and current projections, it is estimated that $10^4$ to $10^7$ interconnected physical qubits are required to break current cryptography standards \cite{mckinsey-24}. However, in the state-of-the-art quantum computers, the number of physical qubits is limited to double digits due to interference, cross-talking, over-heating and challenges in quantum control systems \cite{Pre-18}.

To overcome the aforementioned scalability challenges, multiple quantum processors -- with a limited number of physical qubits -- can be interconnected through a quantum communication network, by adopting a distributed computing approach. This strategy mirrors the classical distributed computing model, where a computation is properly partitioned so that multiple processors can cooperate over a network to solve complex tasks that choke single machine, by leveraging shared data and resources.
Surprisingly, unlike classical systems -- where the computational power scales linearly with the number of interconnected computing modules -- in the quantum domain the computational power can scale exponentially \cite{CacCalTaf-20}, thanks to entanglement. For this reason, both the research community and leading industry players are currently focusing on the development of quantum networks for enabling distributed quantum computing. Notably, IBM has already successfully linked two quantum processors and announced plans to scale this to a seven-QPU (Quantum Processing Unit) network.\cite{CarVaz2024}. 

\subsection{Entanglement-enabled communications}
\label{sec:01.a}

In distributed quantum computing, 
qubits are physically separated across multiple processors. Thus, when a quantum gate must act on remote qubits, namely on non-co-located qubits, specialized communication primitives are required to enable inter-processor operations \cite{CalAmoFer-22}. However, these primitives cannot be borrowed from classical distributed computing model, since the underlying quantum physics demands a fundamentally different paradigm.

To better substantiate the above statement, let us provide the following considerations.\\ 
Unlike classical bits, qubits are subjected to quantum decoherence, 
a phenomenon that irreversibly
scrambles quantum states and therefore the encoded information \cite{CalAmoFer-22}. This kind of quantum noise affects every stage of the quantum distributed computing: processing, storage, and transmission. Accordingly qubits cannot be stored indefinitely, but must be processed within the coherence time. 
Furthermore, differently from classical distributed paradigm, the direct transfer of quantum data from one processor to another is not easy, due to the \textit{no-cloning} theorem, which forbids replicating unknown quantum states.  
As a consequence, the direct transmission of a qubit between processors is inherently risky. Indeed, if a qubit carrying quantum information is lost or corrupted during transmission on the physical channel, its quantum state -- and thus the encoded information-- is irreversibly lost. This makes direct transmission unreliable. 

Thankfully, quantum entanglement, a unique quantum phenomenon where particles share correlated states regardless of the distance, can be exploited
as the key communication resource to avoid the issues arising with the direct transmission of qubits. 
More into details, entangled states, such as Bell pairs, can be distributed in advance through a physical quantum channel\footnote{Since entangled states are communication resources and do not encode information, they are known states. Thus, they are not subjected to the limitations of the no-cloning theorem. As a result, repeated transmission attempts are allowed, until successful distribution is achieved.}. Once established, the entangled pair enables the quantum teleportation protocol \cite{CacCalVan-20}, which leverages the non-local effects activated by the shared entanglement to ``transmit'' the informational qubit without the physical transfer of the particle encoding information. 
These principles form the foundation of the computational paradigms known as \texttt{TeleData} and \texttt{TeleGate}, which respectively allow the transfer of quantum data and quantum gates across distributed quantum systems \cite{CalAmoFer-22}. These \textit{primitives} generalize the concept of moving quantum states across multiple processors in distributed quantum computing, by avoiding the unreliability of direct quantum information transmission. In a nutshell, entangled states, once shared, serve as a virtual link, enabling remote operations to be performed across multiple non-co-located qubits, as they resided within the same quantum processor.

\begin{figure*}
    \centering
    \includegraphics[width=0.8\linewidth]{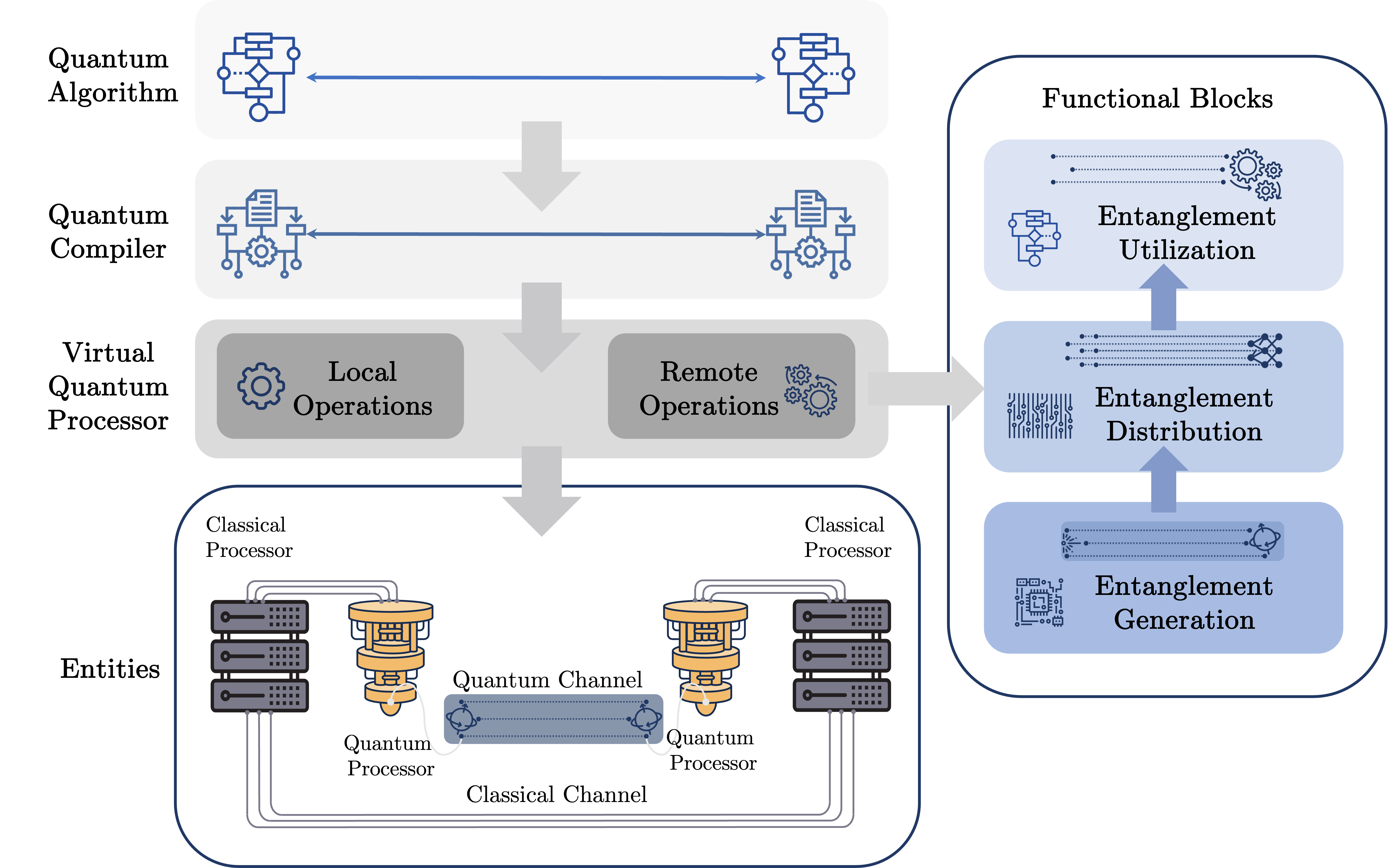}
    \caption{High-level representation of the Distributed Quantum Computing. A crucial in-the-middle component is the compiler, an entity responsible of translating and hardware-agnostic description of the algorithm into a partitioned and fitted version to be executed over the set of individual quantum processors \cite{CuoCalKrs-21,FerCacAmo-21}. As represented in the lowest part of the figure, the set of quantum processing units act as a virtual quantum processor thanks to the interplay of several entities comprising the network infrastructure, represented as quantum communication channels and classical channels, quantum processors and classical processors. Remarkably, the distributed landscape is enabled though three main entanglement-based functional blocks: entanglement generation, entanglement distribution and entanglement utilization.}
    \hrulefill
    \label{fig:01}

\end{figure*}

From this perspective, it emerges that entanglement bridges quantum computing and quantum networks, by unlocking the full potential of distributed quantum computing, as depicted in Fig.~\ref{fig:01}. Specifically, distributed quantum computing relies on three core functional blocks: \textit{entanglement generation}, \textit{entanglement distribution} and \textit{entanglement utilization}, where, entanglement generation and distribution are essential for enabling the virtual links that interconnect remote quantum processors. 

However, entanglement is not solely confined to a functionality reminiscent of the classical physical layer. Its influence extends far beyond. For instance, quantum algorithms leverage highly entangled states to perform computations. This gives birth to an additional functional block, namely, the entanglement utilization block. Furthermore, as the distributed systems scale, entanglement becomes critical in supporting a broad range of functionalities to overcome physical constraints and enable networking functionalities, as deeply detailed in the next sections. 

The remaining part of this work is organized as follows. In Sec.~\ref{sec:02} describes the infrastructural archetypes of quantum networks designed for the distributed quantum computing with a focus on Quantum Data Centres. Then, in Sec.~\ref{sec:03}, we provide a detailed discussion on the operational principles, the physical constraints, and topological challenges of Quantum Data Centres from a communication engineering perspective. In Sec.~\ref{sec:04} future perspectives on the distributed quantum computing landscape are presented with a discussion on the road beyond Quantum Data Centres. Specifically, we discuss the communication requirements for the development of the so-called Quantum Hubs. Finally, Sec.~\ref{sec:05} concludes the paper with open issues and further insights.
\section{Distributed quantum computing: Archetypes}
\label{sec:02}

As pictorially exemplified in Fig.~\ref{fig:01}, in distributed quantum computing architectures, quantum algorithms are executed through coordinated operations across multiple interconnected quantum processors. To this aim, a crucial in-the-middle component is the compiler, responsible of translating a hardware-agnostic description of the algorithm into partitioned and hardware-specific instructions to be executed over the set of individual computing units. 
As represented in the lowest part of the figure, the compiler interacts with the interconnected quantum computing processors, which act as a whole as a virtual quantum processor. And remote quantum operations, namely operations on qubits belonging to different quantum processors, are implemented by exploiting entangled states.
\begin{figure}
    \centering
    \includegraphics[width=1\linewidth]{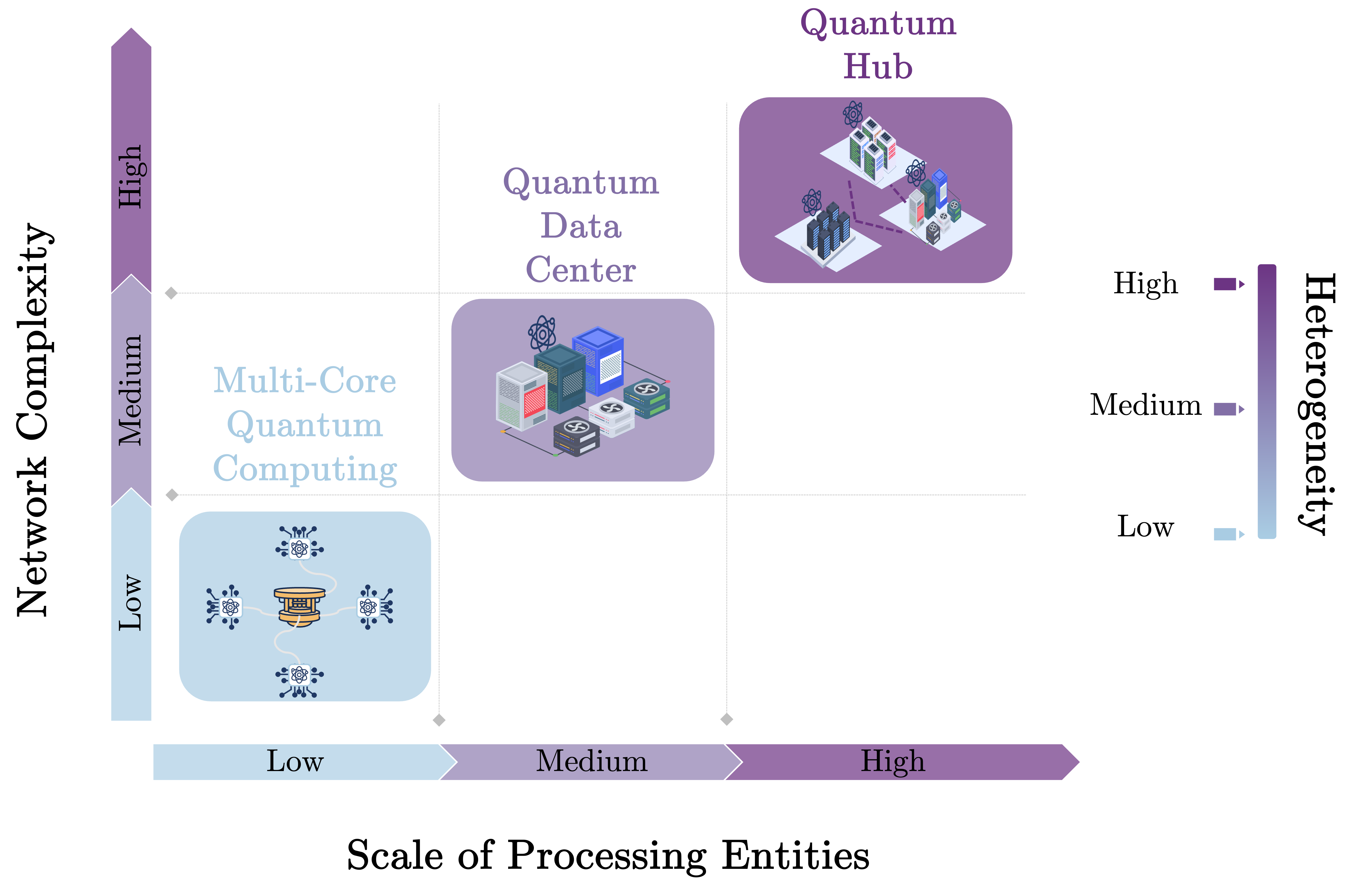}
    \caption{Representation of the Distributed Quantum Computing Archetypes. The x-axis denotes the scale of the interconnected processing entities, the y-axis represents the grade of network complexity for the interconnection of the processing entities, the color-bar denotes the grade of heterogeneity required.  }
    \hrulefill
    \label{fig:02 }
\end{figure}

By enriching with details the abstraction represented in Fig.~\ref{fig:01}, three main archetypes for distributed quantum computing can be recognized \cite{CalAmoFer-22}: Multi-Core quantum computer, \textit{Quantum Data Centres} and Quantum Hub.
As summarized in Fig.~\ref{fig:02 }, the three archetypes differ each others for three main features: the scale of involved processing entities, the complexity of the network infrastructure, enabling the distributed computation and the technology heterogeneity.

Multi-core archetype is characterized by the lowest values across the feature indicators. As detailed in \cite{CalAmoFer-22,RusVinPal-25}, this archetype marks the boundary between single-core quantum computing architectures and distributed quantum computing architectures. In the Multi-Core archetype, multiple quantum processing units are interconnected within a single quantum computer. The most straightforward example of Multi-Core architecture is the deployment of multiple superconducting-based quantum chips within the same cryostat or through advanced couplers \cite{ManCorGam-24,Abu-25}. This archetype exhibits the potential of increasing the number of interconnected physical qubit at a relatively low-cost and efforts, being characterized by low-to-no heterogeneity. Furthermore, Multi-Core architectures exhibit low-network complexity as the network connections can be hard-designed, by leveraging chip-scale interconnections. This, in turn, implies that the number of interconnected cores is inevitably bounded by the rack/refrigerator volume. Thus, the number of physical qubits that can be clustered together is limited as well. Accordingly, the network functionalities are also relatively simple\cite{CalAmoFer-22,EscRacRod-23}.

As represented in Fig.~\ref{fig:02 }, the Quantum Hub, namely the distributed quantum computing configuration interconnecting multiple diverse Quantum Data Centres, is the most complex archetype in our classification, being characterized by the highest grade of network complexity, heterogeneity and scale of the interconnected entities. Indeed, significant
heterogeneity may arise, since different Quantum Data Centres are likely owned by different operators or rely on different hardware platforms. Furthermore,
the interconnection of geographically-distributed centres requires a wide-scale network infrastructure, likely enabled by the
Quantum Internet \cite{CalAmoFer-22,Kim-08,rfc9340}.
In this context, it is clear that network interoperability across platforms should be supported, while ensuring a reliable management of entanglement generation and distribution at large scale. This, in turn, requires seamless integration with the classical Internet infrastructure \cite{CacIllKou-22,CacIllCal-23}.

While current distributed quantum systems remain in early development, the Quantum Hub archetype emerges as a more advanced stage of the technological evolution.


\subsection{Multi-Computer Archetype: Quantum Data Centre }

Quantum Data Centres represent the bridge between the small-scale distributed quantum computing systems and the large-scale, high-complexity archetype \cite{LiuJia-24,LiuHanJia-23-1,AghAinAle-25}. In this intermediate configuration, multiple quantum computers -- each comprising of multiple quantum cores -- are located in the same controlled environment (e.g. a dedicated facility) and interconnected via a Quantum Local Area Network (QLAN) \cite{MazCalCac-25,MazCalCac-24-QCNC,MazZhaChu-24}. 

Accordingly, the physical distance between remote qubits in the multi-computer archetype ranges from room-scale to building-wide. Thus it is significantly larger than in Multi-Core systems but smaller than in the quantum hub archetype. Moreover, the number of physical qubits that can be clustered together is bounded
by the number of computers that can be interconnected within the same data center. This, in turn, is highly influenced by the QLAN features as detailed in the next sections. 
These extended distances introduce more complex challenges with respect to the Multi-Core archetype, such as the transmission impairments on quantum carriers, or higher classical control latency. Furthermore, the QLAN must support advanced network functionalities, for coordinating distributed quantum operations across multiple nodes, including entanglement routing and error mitigation. In this archetype a medium degree of heterogeneity is expected. This means that different types of quantum hardware (e.g. superconducting qubits, photonic qubits, atoms in cavities) might coexist with a specialized quantum transducer needed to convert quantum information between different supports efficiently, as better detailed in the next section.

As aforementioned in the introduction, the design and development of Quantum Data Centres represent the next quantum engineering challenge, as already acknowledged by the major stakeholders and tech players in the related areas\cite{Cisco-24,ManCorGam-24, ibm2024}. 
For this, the remaining part of the paper focuses on the architectural and network operational challenges of Quantum Data Centres from a communication engineering perspective. Before concluding this section, it is worthwhile to mention that given the inherent complexities of the Quantum Data Centre archetype, we envision both classical and quantum Machine Learning (ML) techniques to play a relevant role in its evolution. Indeed classical and quantum ML techniques can be exploited for managing and optimizing critical processes, including quantum control, error mitigation, and entanglement distillation strategies\cite{NeuWez-22,BenLloSac-19,ChiSim-23,PhiChi-22,ChiHar-23,AdeKum-25,LocCarMul-23}.

\section{Quantum Data Centres: constraints on implementation and physical topology}
\label{sec:03}

As discussed, entanglement serves as fundamental resource enabling quantum operations between qubits distributed across different quantum computers within a Quantum Data Centre. However, its implications extend far beyond this functionality. Indeed, entanglement revolutionizes the very same concept of a communication network, empowering an entirely new concept of connectivity, transcending the classical networking limitations\cite{IllCalMan-22, CalAmoFer-22, MazCalCac-25, MazZhaChu-24, CheIllCac-24}.
Consequently, the ability to decouple the physical QLAN network topology from the actual capacity to exchange quantum information in a Quantum Data Centre, via the entanglement-activated connectivity, introduces novel and largely unexplored network design paradigms\cite{IllCalVis-23, CalAmoFer-22, MazZhaChu-24, MazCalCac-25}.
More into details, QLAN deployments require resource-intensive, costly, and sophisticated setups \cite{DavCacCal-2024, CalDavHan-25, MazCalCac-25}. This, in turn, implies, at least in the mid-term time horizon, that:
\begin{itemize}
    \item the QLAN physical topology significantly diverges from the highly connected, densely populated, and highly dynamic classical data centre topologies \cite{MazCalCac-25};
    \item the ability to generate, distribute, and manipulate entangled states is confined into specialized nodes. In fact, technology maturity prevents homogeneous node functionalities\cite{RuiWalDur-25, DavCacCal-2024, DavZhaChu-24}. 
    \end{itemize}
These constraints necessitate a hierarchical QLAN design where specialized nodes, referred to as the \textbf{\textit{orchestrator}} nodes, are responsible for the entanglement generation and distribution \cite{DavCacCal-24, CheIllCac-24, MazCalCac-25}, according to different strategies \cite{CacIllVis-24}. 
Thus, as extensively discussed later in this section, in the \textit{orchestrator} node is concentrated the complexity, by handling the most resource-intensive network operations. While the remaining network nodes are simple, by exhibiting limited capabilities.
\subsection{ Quantum Transduction}

Quantum Data Centres inherently rely on heterogeneous  qubit platforms, since no single qubit technology currently meets all the requirements needed for quantum computation and communication, as detailed in the following.

\textit{Superconducting technology} stands out as a leading platform for quantum computations for its key advantages.
In fact, fast, high-fidelity gates can be realized via this hardware platform, by ensuring high circuit scalability. Furthermore, the superconducting technology is characterized by efficient control and readout mechanisms \cite{KjaSchBra-20, Wen-17, OfePetRei-16}. 
However, these benefits come at the price of superconducting qubits operating at cryogenic temperatures (few tens of mK), which makes them very susceptible to thermal noise and decoherence. 

With this in mind, for connecting superconducting nodes in a Quantum Data Centre for distributed quantum computation, in principle cryogenic waveguides \cite{XiaZhaLia-17, KurPechRoy-19, ManStoKur-20} can be utilized.
However, cryogenic cables present significant practical limitations. Indeed, they are characterized by very expensive manufacturing costs and non-trivial installation requirements. Furthermore, they introduce severe communication constraints, such as very limited communication range and inflexible network topology. This, in turn, results in huge difficulties in reconfiguring the network architecture. Consequently, purely cryogenic interconnection strategies are not viable for scaling distributed quantum computing systems.

\begin{figure}
    \centering
    \includegraphics[width=0.7\linewidth]{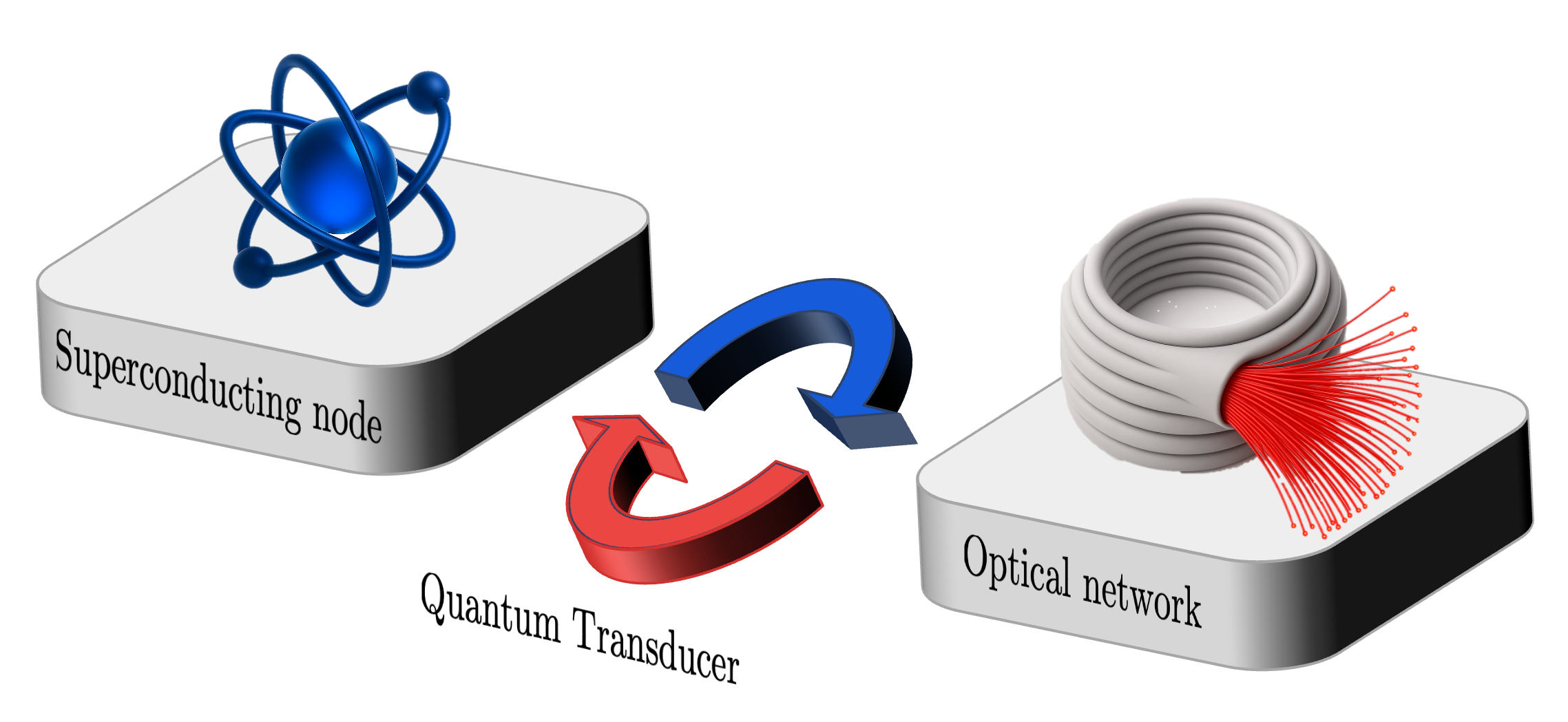}
    \caption{Schematic representation of the role of a quantum transducer as an interface between superconducting nodes and optical network.}
    \hrulefill
    \label{fig:03 }
\end{figure}

\begin{table}[ht!]
\centering
\renewcommand{\arraystretch}{1.5} 
\caption{Comparison between cryogenic cables and optical fibers for quantum communication}
\begin{tabular}{|c|c|c|}
\hline
 & \textbf{Cryogenic Cables} & \textbf{Optical Fibers} \\
\hline
\textbf{Quantum Transduction} & Not Required & Required \\
\hline
\textbf{Communication Range} & Limited distances & Long distances \\
\hline
\textbf{Flexibility} & Limited & Flexible \\
\hline
\textbf{Infrastructure} & Expensive waveguides & Existing fiber networks \\
\hline
\end{tabular}
\label{tab:01}
\end{table}
An alternative solution for interconnecting superconducting processors leverages an optical fiber network for entanglement distribution. 
Indeed, \textit{quantum optics technology} offers several advantages: weak interaction with the environment (low decoherence), high-fidelity transmission, easy control with standard optical components and high-speed operations \cite{RenXuYon-17,CacCalTaf-20}.
Therefore, optical photons, also referred to as \textit{flying qubits}, are widely recognized as entanglement carriers.

Compared to cryogenic-based interconnections, fiber-based solutions offer key several assets not limited to the significantly lower infrastructure costs. Specifically, they enable long-range communications, and accommodate diverse network topologies, by allowing dynamic reconfigurations of the network architecture. In addition, a compelling advantage of optical quantum networks is their ability to leverage existing telcom-fiber infrastructures. In fact, recent results show that entanglement distribution can be achieved by utilizing lit fibers, without additional costs, by eliminating the need of dedicated quantum channels, such as dark fibers. Clearly, for a seamless integration with classical networks, the coexistence of classical and quantum signals -- sharing the same fiber -- has to maintain quantum integrity via noise suppression techniques \cite{JorFeiChe-24}. The above comparison is summarized\footnote{We kindly refer the interested reader to \cite{CalDavHan-25} for a thorough and comprehensive discussion on quantum transduction from a communication perspective} in Tab.\ref{tab:01}.

However, the crucial obstacle for interconnecting superconducting nodes with an optical QLAN network stems from the huge frequency gap between optical photons (that typically work at about hundred of THz) and superconducting qubits (that work at a few GHz). 
This five-order-of-magnitude gap needs \textit{Quantum Transduction} \cite{CacCalTaf-20, LauSinBar-20}, i.e. the process of converting one type of qubit to another, by enabling the interaction of different qubit platforms \cite{CalDavHan-25}.
A quantum transducer, namely the network component performing quantum transduction, constitutes a mandatory interface converting a superconducting qubit within a network node into a flying qubit that travels through optical channels and vice versa \cite{CalDavHan-25, DavCacCal-24}, as schematically depicted in Fig.~\ref{fig:03 }.
However, as clarified in the next paragraph, enabling an effective quantum transduction, by preserving the quantum states, is still an open problem for the current technology maturity.
Therefore, quantum transduction remains the main limitation in utilizing optical fibers for interconnecting superconducting nodes in a Quantum Data Centre.

\subsubsection{Direct vs Entanglement-Generation Transduction}
\label{sec:03.1.a}
For an in-depth treatise, we refer the reader to \cite{CalDavHan-25}. The same hardware to implement a quantum transducer can be exploited for performing two distinct functionalities: 
\begin{itemize}
    \item \textit{Direct Quantum Transduction} (DQT): the transducer enables the direct conversion of one type of qubit to another, regardless if the state of the qubit encodes information or quantum correlation for entanglement distribution. In this last case, the qubit is referred to as entanglement qubit (ebit). For both informational qubits and ebits, the frequency conversion can be performed in two directions:
    \begin{itemize}
        \item[-] \textit{up-conversion}: the quantum transducer converts the state of a superconducting qubit operating at microwave frequency $\omega_m$ into a degree of freedom of an optical photon operating at frequency $\omega_o$.
        \item[-] \textit{down-conversion}: the quantum transducer converts the state of an optical photon, operating at optical frequency $\omega_o$, into the state of a superconducting qubit, operating at microwave frequency $\omega_m$.%
    \end{itemize}
    \item \textit{Entanglement Generation Transduction} (EGT): the transducer enables the entanglement generation.
\end{itemize}

\begin{figure*}
\hfill

\begin{subfigure}{\linewidth}
    \includegraphics[width=\linewidth]{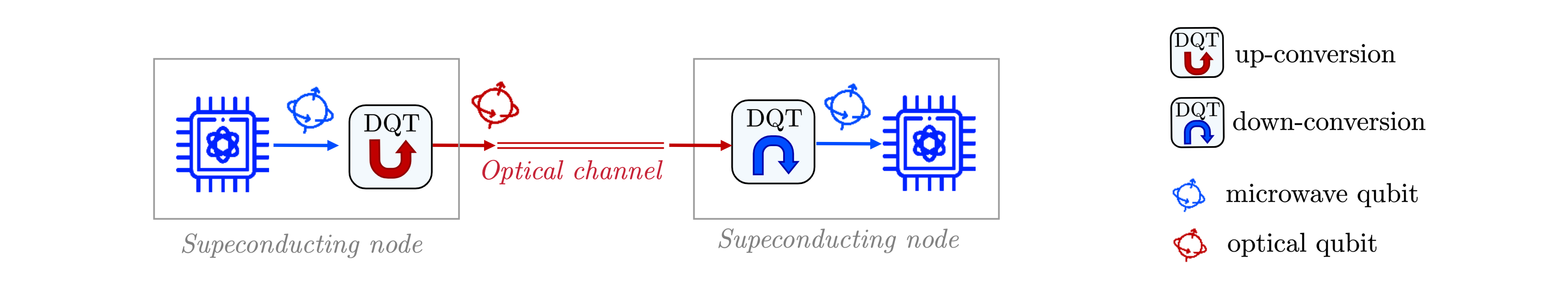}
    \caption{DQT on qubit.}
    \label{fig:DQTqubit}
\end{subfigure}
\hfill
\begin{subfigure}{\linewidth}
    \includegraphics[width=\linewidth]{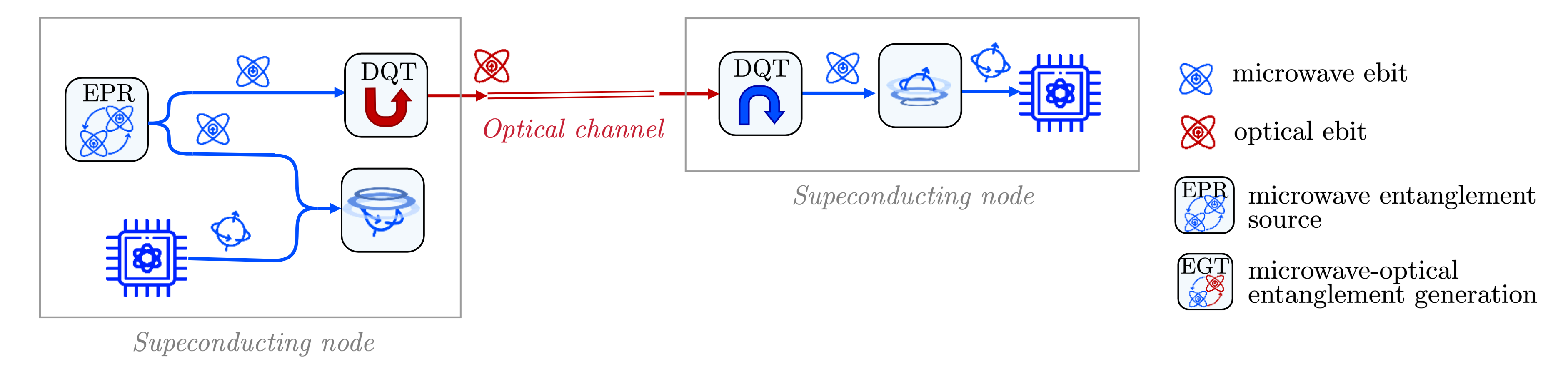}
    \caption{DQT on ebit.}
    \label{fig:DQTebit}
\end{subfigure}

\hfill
\begin{subfigure}{\linewidth}
    \includegraphics[width=\linewidth]{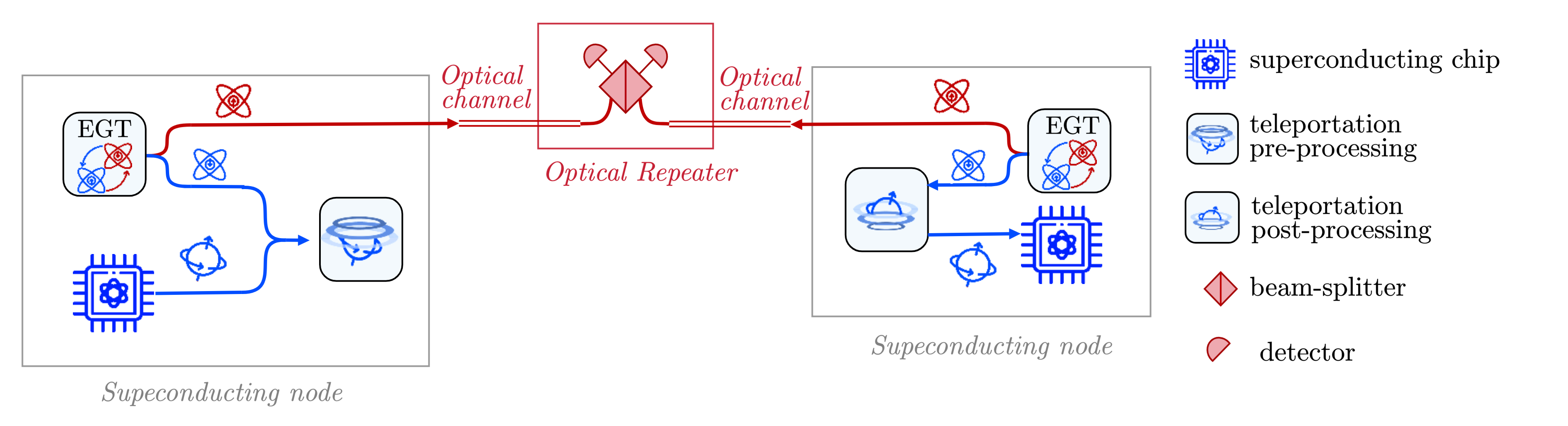}
    \caption{EGT coupled with entanglement swapping.}
    \label{fig:EGT}
\end{subfigure}
\caption{Different transducer-based archetypes for interconnecting superconducting processors. Qubits and ebits at microwave (optical) frequency are depicted in blu (red).}
    \hrulefill

\label{fig:04}
    \end{figure*}

Different transducer-based archetypes, which exploit the different functionalities of quantum transduction, can be envisioned for interconnecting superconducting processors, as schematically summarized in Fig.~\ref{fig:04}. It is worthwhile to note that when DQT acts on informational qubits, it enables the direct transmission of quantum information. However, Section~\ref{sec:01.a} highlighted the fundamental limitations of direct transmission in distributed quantum computation architectures, arising from the no-cloning theorem and quantum measurement postulate \cite{CacCalVan-20}. Indeed, the challenges in DQT on informational qubits are even more critical than standard direct quantum transmission, since transduction errors compound with transmission losses. In fact, the quantum transduction should be able to preserve the encoded quantum information in the conversion, that constitutes a not trivial task.
Consequently, distributed quantum computing architectures must rely on entanglement-mediated approaches, such as DQT on ebits or EGT.

DQT on ebits is exploited for entanglement distribution. Specifically, the entanglement resource is locally generated and then distributed to the remote node via quantum transduction. Once the entanglement is successfully generated and distributed, remote operations can be executed accordingly to the TeleData and TeleGate paradigms as in Fig.~\ref{fig:DQTebit}.
The main advantage of DQT on ebits lies in the possibility of entanglement re-generation \cite{IllCalMan-22}. Indeed,
if the ebits are lost in the conversions (up- at source or down- at destination), they can be regenerated without restrictions,
being a communication resource rather than an information. The re-generation process can be performed until the entanglement is successfully distributed between remote nodes \cite{CalDavHan-25}.

In the EGT paradigm, the transducer generates hybrid entanglement between different frequencies domains \cite{CalDavHan-25, DavZhaChu-24, DavCalWan-23, MeeLakWoo-24, SahQiuHea-23}, i.e., entanglement between a microwave photon and an optical photon in the form of a 2-level Fock state\footnote{The approximation of a two-level Fock-state is satisfied with some hypothesis regarding the physical interaction underlying transduction. We refer the interested reader to \cite{CalDavHan-25} and \cite{DavCacCal-24}.}:
\begin{align}
   \label{eq:EGT}
    \ket{\Psi_{m,o}} = \alpha\ket{0_m 1_o}+\beta\ket{1_m 0_o},
\end{align}
where the subscripts $m$ and $o$ refer to the frequency domain of each ebit, namely microwave and optical, and $\alpha$ and $\beta$ depends on the transducer hardware \cite{DavCacCal-24,DavCacCal-2024}. EGT can be exploited to implement the Duan–Lukin–Cirac–Zoller protocol \cite{DuaLukCir-01}, represented in Fig.~\ref{fig:EGT}.
Accordingly, the optical ebits generated in each node with EGT travel through the optical fibers to reach an optical \textit{repeater node}. The repeater is equipped with a 50/50 beam splitter followed by two detectors. The click of one detector denotes the presence of an optical photon. In this configuration when a detector clicks, it is impossible to identify from which transducer the optical photon comes from, or, equivalently, it is impossible to distinguish whether a microwave photon is present at source or at destination. In other words, the overall setup is unable to distinguish the so-called \textit{which-path} information, that, by oversimplifying, is related to the knowledge about the path a photon takes through a multipath optical system, such as a beam splitter.

This results into the generation of \textit{path-entanglement} \cite{MontVivCap-15} between the microwave photons at the source and at the destination:
\begin{equation}
    \label{eq:EGTswap}
    \ket{\Psi_{m,m}^{s,d}}=\frac{1}{\sqrt{2}}(\ket{0_m^s1_m^d}+\ket{1_m^s0_m^d}),
\end{equation}
where the up-scrips $s$ and $d$ denote the ``location" of each ebit, namely at source and destination. 
To summarize, the shared entangled pair in \eqref{eq:EGTswap} between two microwave photons at the source and destination is obtained from one entangled state between an optical photon and a microwave photon at the source and one entangled state between an optical photon and a microwave photon at the destination. Consequently, the overall result is reminiscent of \textit{entanglement swapping}\cite{CalDavHan-25, DavCacCal-2024}. 

This archetype for entanglement distribution exploiting two EGTs and entanglement swapping allows to herald entanglement between remote superconducting devices, without destroying it. Once the distributed entanglement is heralded, it can be leveraged for remote operations.

\begin{figure}[htp]
    \centering
    \begin{subfigure}[b]{0.45\textwidth}
        \centering
        \includegraphics[width=\textwidth]{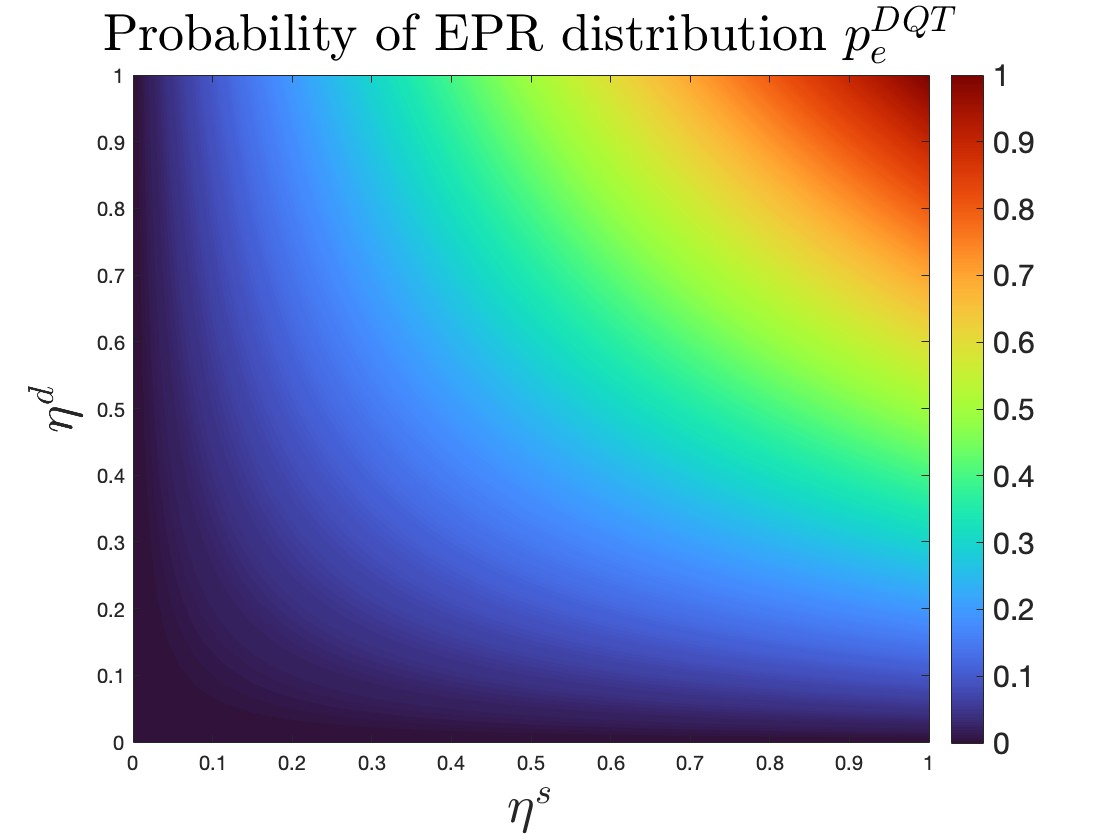}
        \subcaption{Probability of EPR distribution with DQT.}
        \label{fig:simulation1}
    \end{subfigure}
    \hfill
    \begin{subfigure}[b]{0.45\textwidth}
        \centering
        \includegraphics[width=\textwidth]{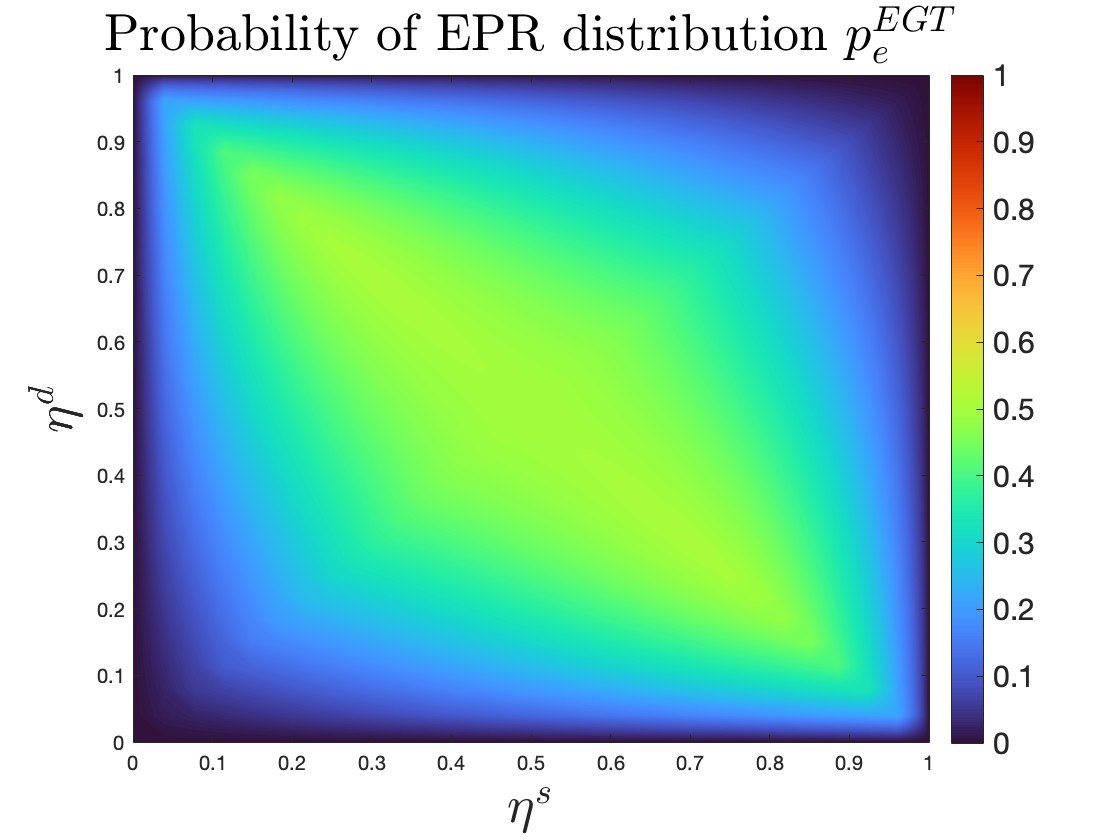}
        \subcaption{Probability of EPR distribution with EGT.}
        \label{fig:simulation2}
    \end{subfigure}
    
    \caption{Probability of EPR distribution $p_e$ in DQT on ebits and in EGT. The superscripts $^s$ and $^d$ indicates that the conversion efficiency refers to the source or destination, respectively.}
    \label{fig:05}
\end{figure}

The key advantage of EGT over DQT on ebits is to relax the constraints on the conversion efficiency $\eta$ -- the main hardware parameter characterizing the transducer performance -- for having a non-null entanglement distribution probability, as depicted in Fig.~\ref{fig:05}. 

The described framework can be adopted in a Quantum Data Centre scenario, where -- as aforementioned -- a specialized node, the orchestrator, is responsible for entanglement generation and distribution to the other nodes \cite{DavCacCal-24, DavCacCal-2024}. This generalization can also include in the picture the possibility of generating not only bipartite entangled states but also multipartite entangled states, allowing new and astonishing network functionalities. This has been deeply discussed from a quantum transduction perspective in \cite{DavCacCal-24, DavCacCal-2024}.

\subsection{QLAN architecture: physical vs artificial topology}
\label{sec:03.2}

As aforementioned, in the Quantum Data Centre archetype, the quantum processors are interconnected via a quantum local area network (QLAN). Densely physical-connected QLAN topologies are not practical in a short term. The rationale is that the deployed architectures have to take into account the high cost of data buses (economically and in terms of quantum fidelity) limiting both the size of the clusters and the use of connections for processing. Indeed, as detailed in the previous section, the connections in such controlled environments must either resort to crycables (expensive and necessarily limited in length, for handling the decoherence effects) or resort to quantum transduction, which is still extremely inefficient. Thus, as clarified at the beginning of Section~\ref{sec:03}, the maturity level of quantum technology leads to two main characteristics of the QLAN. First, the physical topology is sparse. Second, there is a hierarchical structure within the QLAN, where specialized nodes, 
the orchestrator nodes, are responsible for the entanglement generation and distribution. Consequently, the network logic is centralized at the orchestrator node, while the remaining nodes, referred in the following as client nodes, are lightweight and simple \cite{MazCalCac-24-QCNC, MazCalCac-25}. Although this design fundamentally departures from classical network paradigms, the quantum-specific constraints, including decoherence timescales and hardware complexity, make this centralized approach viable. 
\begin{figure}[t!]
    \centering
    \includegraphics[width=1\textwidth]{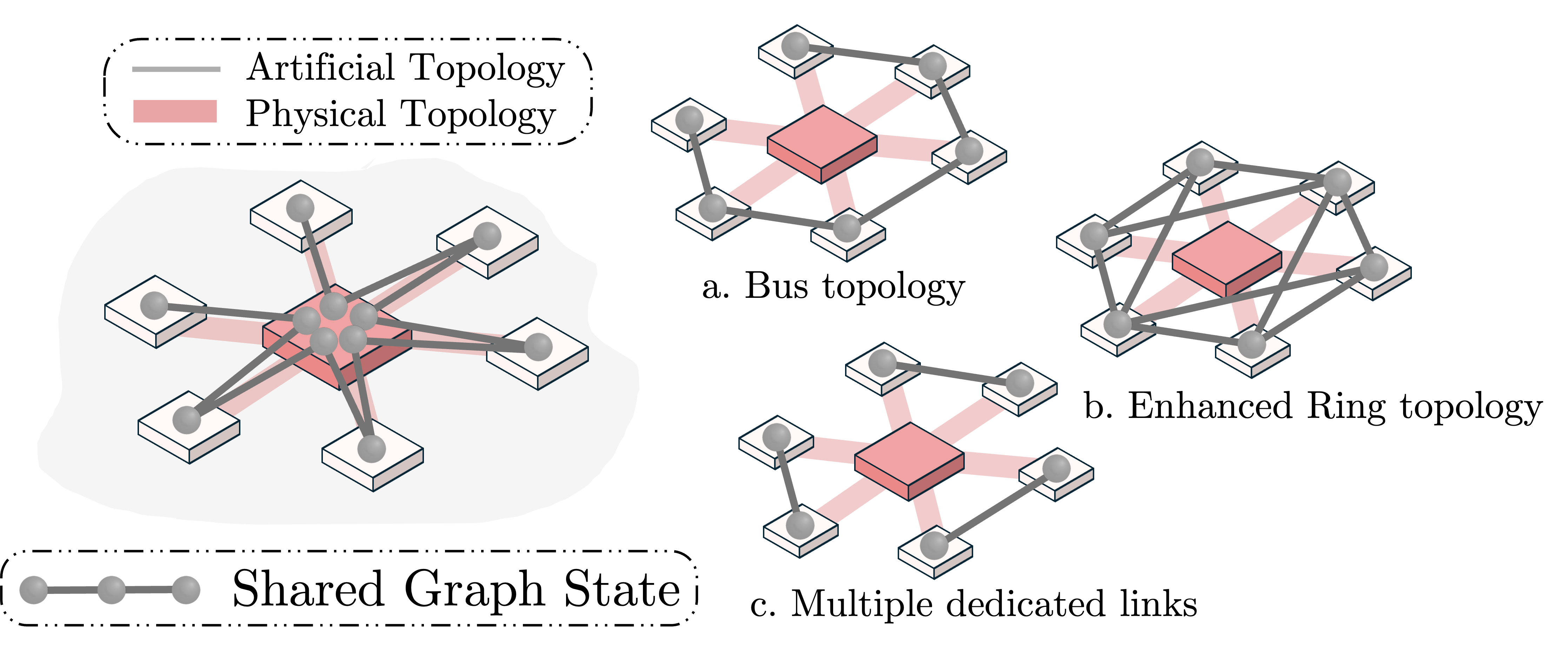}
    \caption{Physical star topology of a QLAN with $6$ client nodes and a single orchestrator (left). Examples of different artificial topologies obtainable by proper manipulations of the shared multipartite entangled state (right).}
    \hrulefill
    \label{fig:06}
\end{figure}
The simplest configuration of the described QLAN architecture, with a single orchestrator connected via a physical star topology to the clients, is depicted in Fig.~\ref{fig:06}. 

A sparse QLAN physical connectivity inherently restricts the communication capabilities between the QLAN nodes. For overcoming these limitations, we cannot borrow well-established approaches from classical LANs \cite{MazCalCac-25}, where topological constraints are overcome through the upper layers of the protocol stack, at the price of communication overhead and information duplication. This impossibility is due to the unconventional
quantum peculiarities, ranging from stringent coherence
times to quantum mechanics postulates and phenomena,
such as quantum measurement and the no-cloning theorem.
Furthermore, the design of a protocol suite for quantum networks
is still at its infancy \cite{CacIllCal-23}, and thus the functionalities of
“quantum” upper-layers are yet to be defined.

Fortunately, the communication limitations induced by
the physical QLAN topology can be overcome, by relying
on quantum entanglement, which 
enables a fundamentally new form of connectivity, referred to as \textit{\textbf{entanglement-based connectivity}} \cite{IllCalMan-22, MazCalCac-24-QCNC, MazCalCac-25}. 
Specifically, by leveraging multipartite entanglement, it is possible to establish a virtual topology -- referred to as \textit{artificial topology} -- overlaying the physical one. 
The artificial topologies activated by entanglement can significantly differ from the underlying physical topology, and they can -- if properly engineered -- effectively mitigate the constraints imposed by the QLAN architecture.

Indeed, as depicted in Fig.~\ref{fig:06}, the required artificial topology can be engineered to support a wide range of quantum communication patterns.
However, the choice and the design of the multipartite entangled state to be generated and distributed is highly non-trivial and has great influence on the enabled network functionalities \cite{IllCalVis-23, IllCalVis-23, CheIllCac-24}. As instance, it is well known that by sharing a GHZ state, a single Bell state can be extracted, on-demand, whenever needed, between any pair of network nodes sharing part of the entangled state \cite{IllCalVis-23}. This has the advantage to dynamically adapt to the communication demands, without predetermining the identities of the involved nodes\footnote{Conversely, when a Bell state is shared between two parties, the identities of nodes involved in the communication has been fixed a-priori.}.
Although GHZ states present interesting properties compared to the use of bipartite entangled states, they are limited by the extraction of a single Bell state, even if the number of qubits of the overall state grows arbitrarily. At the same time, GHZ states are characterized by unitary persistency \cite{IllCacMan-21}, i.e., the minimum number of qubits that need to be
measured to guarantee that the resulting state is separable is one. Thus, they are highly sensitive to errors: losses of even one of the particles causes the entire state to lose coherence. 

Accordingly, the choice of the multipartite state to be distributed in the network involves fundamental compromises between three main factors: the number of Bell pairs that can be extracted -- after manipulations of the initial state -- persistency and robustness to quantum noise. Thus the degrees of freedom for the design can be resumed as follows \cite{MazCalCac-25}: 
\begin{itemize}
    \item[I)] The "\textit{\textbf{type}}" of quantum state -- its properties and entangled structure -- to be generated, distributed and engineered.
    \item[II)] The \textit{\textbf{number of qubits}} of the multipartite entangled state, which is not necessarily equal to the number of clients in the quantum network.
\end{itemize}

\begin{table}[b!]
\centering
    \begin{tabular}{|c c|}
        \hline
        \hline
        \textbf{Pauli measurements on $\ket{G}$} & \textbf{Corresponding graph operations} \\
        \hline
            
        $P_{z,\pm}^{(a)} \ket{G}  = \ket{z,\pm}^{(a)} \otimes \underbrace{U_{z,\pm}^{(a)}\ket{G - a}}_{\ket{\tilde{G}_z}}$ & $\tilde{G}_z = G - a$ \\
            
        $P_{y,\pm}^{(a)} \ket{G}  = \ket{y,\pm}^{(a)} \otimes \underbrace{U_{y,\pm}^{(a)}\ket{\tau_a(G)-a}}_{\ket{\tilde{G}_y}}$ & $\tilde{G}_y = \tau_a(G) - a$ \\
            
        $P_{x,\pm}^{(a)} \ket{G}  = \ket{x,\pm}^{(a)} \otimes \underbrace{U_{x,\pm}^{(a)}\ket{\tau_{b_0}\left(\tau_{a}\left(\tau_{b_0}(G)\right) - a\right) }}_{\ket{\tilde{G}_x}}$ & $\tilde{G}_x = \tau_{b_0}\big( \tau_{a} (\tau_{b_0}(G)) - a\big)$ \\
            
        \hline
        \hline
    \end{tabular}
    \caption{Table of single-qubit Pauli measurements with projectors $P_{i,\pm}^{(a)}, \, i \in \{x,y,z\}$ acting on a generic qubit $a$ of the graph state $\ket{G}$ (with outcome $\pm1$) and corresponding graph operations, in terms of vertex deletions and local complementations $\tau(\cdot)$. Table reproduced from \cite{MazZhaChu-24}.}
    \label{tab:Pauli_Meas}
\end{table}

\subsubsection{Graph states as network resources}

\begin{figure}[t!]
    \centering
    \begin{subfigure}[b]{0.33\textwidth}
        \centering
        \includegraphics[width=1\textwidth]{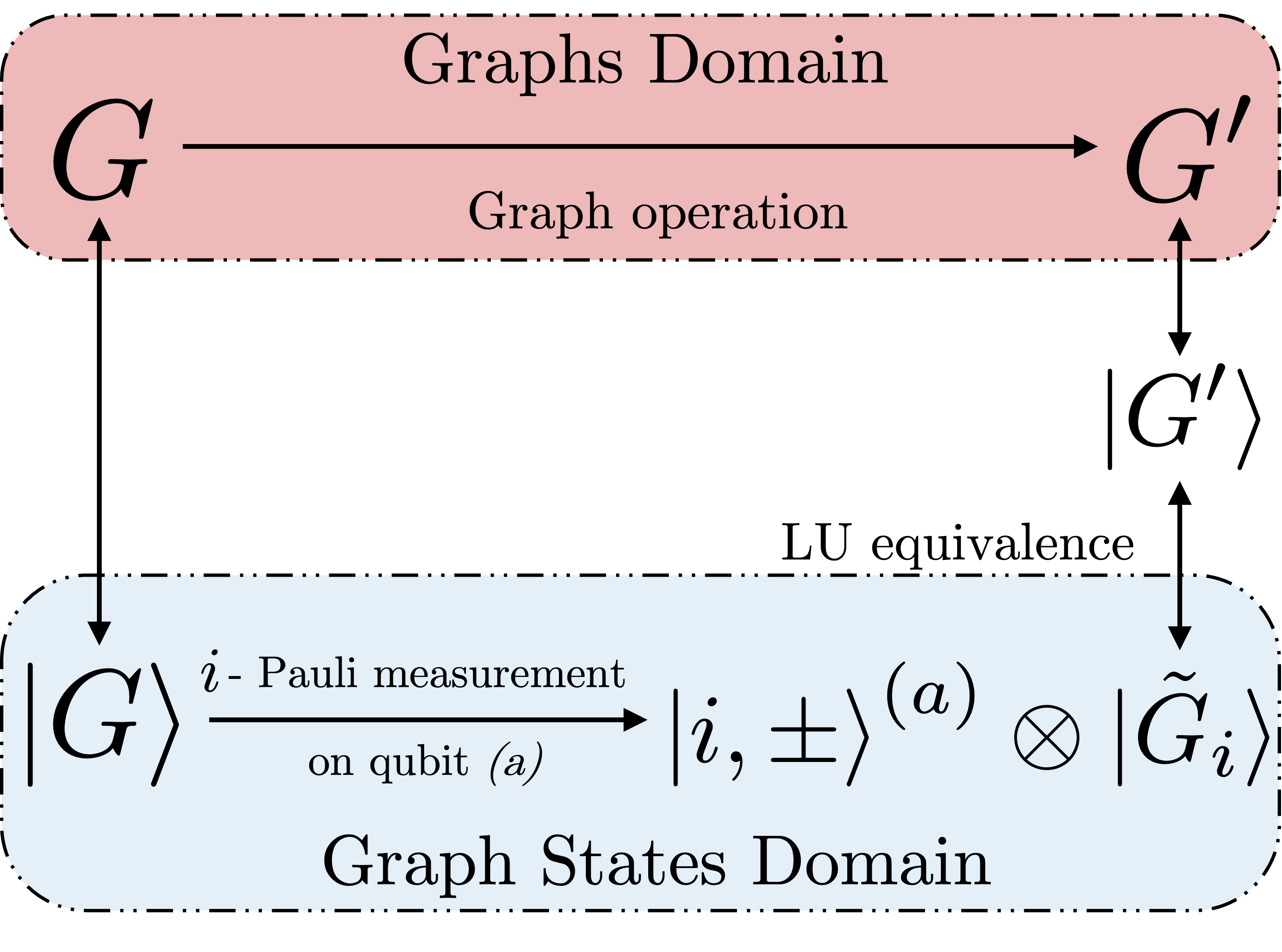}        \subcaption{Correspondence between graph domain and graph states domain.}
        \label{fig:07a}
    \end{subfigure}
    \hfill
    \begin{subfigure}[b]{0.65\textwidth}
        \centering
        \includegraphics[width=1\textwidth]{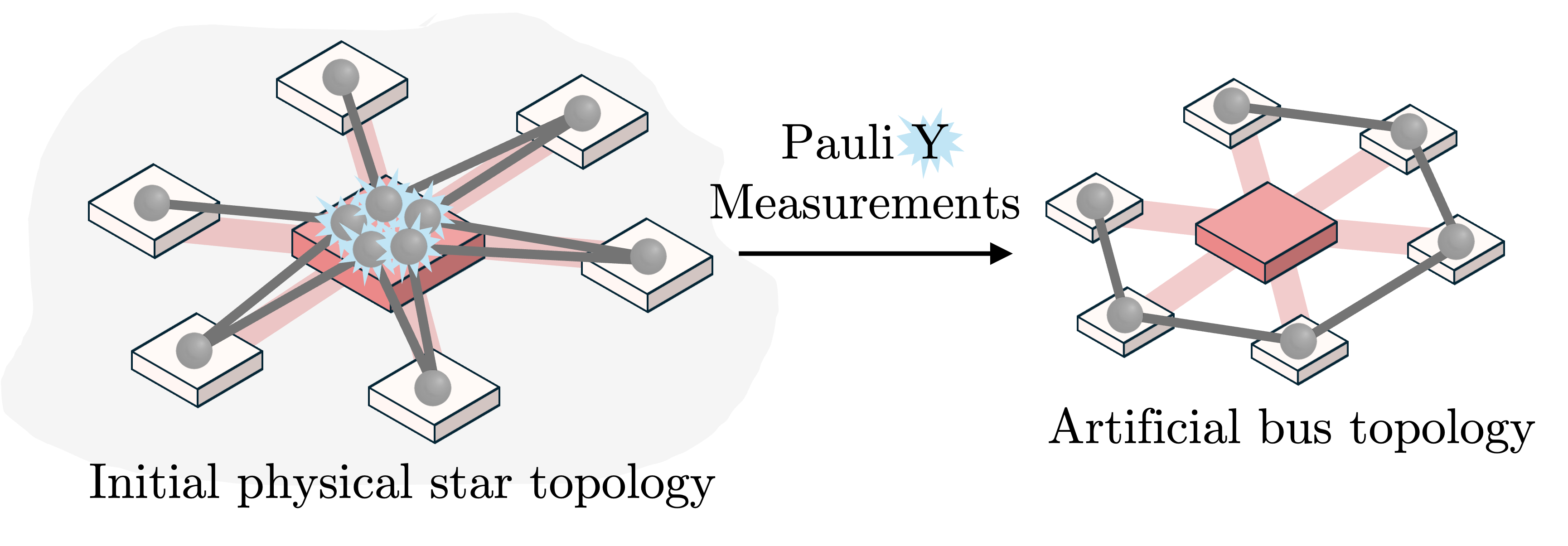}        \subcaption{Correspondence between (initial) physical star topology and artificial bus topology, according to the application of Pauli-Y measurements on the qubits retained at the orchestrator.}
        \label{fig:07b}
    \end{subfigure}
    \caption{Schematic representation of the correspondence between graph states measurements and the associated graph operations (a). Application of Pauli-Y measurements for the engineering of the artificial topology of the network in a $6$ clients QLAN topology (b).}
    \hrulefill
    \label{fig:07}
\end{figure}

Stemming from the above discussion, graph states are particularly interesting candidates, due to their unique entanglement properties, noise robustness and intrinsic resilience to particle losses \cite{HeiEisBri-04, HeiDurEis-06, RuiDur-24, RuiWalDur-25, AigRuiDur-25}.
Interestingly, graph states have a 
straightforward representation of entanglement interactions in the form of a graph. In other words, each graph state $\ket{G}$ has its own associated graph $G=(V,E)$, and can be written in the following form:
\begin{equation}
    \label{eq:graph_state}
    \ket{G} = \prod_{\{a,b\} \in E} \texttt{CZ}_{ab}\ket{+}^{\otimes n},
\end{equation}
where $\ket{+}= \tfrac{1}{\sqrt{2}} (\ket{0} + \ket{1})$ and where the controlled-Z gate ($\mathtt{CZ}$) represents the entangling operation between qubits $a$ and $b$, corresponding to an edge in the associated graph $G=(V,E)$. 
One of the most interesting properties of graph states is the possibility to engineer the state structure, by wisely applying local quantum operations on some qubits. More into details, the application of single-qubit Pauli measurements -- denoted by the projection operator $P_i$ with $i \in \{x,y,z\}$ -- map into specific graph operations on the associated graph\cite{HeiEisBri-04, HeiDurEis-06}. As explicitly reported in Tab~\ref{tab:Pauli_Meas} and visually represented in Fig.~\ref{fig:07a}, the graph state obtained after each Pauli measurement is equivalent -- up to the application of unitary operators $U_{i,\pm}$ -- to a new graph state, whose associated graph is obtained through vertex deletions and local complementations $\tau(\cdot)$ on the original graph $G$ \cite{HeiDurEis-06, MazCalCac-25, MazZhaChu-24}.

The simplest graph state is the $n$-qubits \textit{linear} graph state, whose associated graph is a linear interconnection of vertices and its expression reduces to: $\ket{L} = \prod_{i=1}^{n-1} \texttt{CZ}_{i,i+1}\ket{+}^{\otimes n}$.
Such a simple graph state is particularly interesting in the context of quantum communication and computation \cite{ThoRusRem-24, BarBirBom-23}, since it can be used as building block for more complex target states \cite{ShaSha-23}, according to proper manipulations and merging of multiple instances, as experimentally proven in controlled environments \cite{ButBarEco-17, HilLeoEis-23, LeeJeo-23}.

\subsubsection{From physical to artificial topologies}

Accordingly to the above, graph states emerge as promising resource for engineering the QLAN topologies, due to their unique entanglement properties and operational flexibility\cite{MazCalCac-24-QCNC, MazZhaChu-24, MazCalCac-25}. The centralized QLAN model, depicted in Fig.~\ref{fig:06}, allows the orchestrator node to generate and distribute graph states and linear graph states, while minimizing client-side operations and associated communication overhead. To this aim, it retains for itself part of the qubits of the generated graph state while distributing the remaining ones to the client nodes. With this strategy, the orchestrator node can engineer the artificial topology solely through the use of its locally held qubits and by employing Local Operations and Classical Communication (LOCC). Thus, this approach allows the dynamic reconfiguration of the QLAN topology, without requiring distributed quantum operations at the clients, besides the correction unitaries after the Pauli measurements, as indicated in Tab~\ref{tab:Pauli_Meas}. 

An example of this topology engineering is illustrated in Fig.~\ref{fig:06} (right) and Fig.~\ref{fig:07b}. In this scenario, the orchestrator initially generates a linear graph state and distributes part of it, enabling the establishment of entangled links between otherwise unconnected clients. By performing Pauli-Y measurements on its retained qubits (Fig.~\ref{fig:07b}), the orchestrator effectively transforms the entangled structure from a star configuration into a bus topology, sequentially linking the clients through entangled links. 

The orchestrator can leverage the inherent simplicity of linear graph states as building block for synthesizing more sophisticated entangled states, as recently shown in \cite{MazCalCac-25}. This approach enables the engineering of denser QLAN artificial topologies. 
This, in turn, implies that the orchestrator gains 
enhanced capabilities to tailor the connectivity patterns, allowing for more versatile and resource-efficient quantum computation architectures.

\begin{figure}
    \centering
    \begin{subfigure}[b]{0.48\textwidth}
        \centering
        \includegraphics[width=1\textwidth]{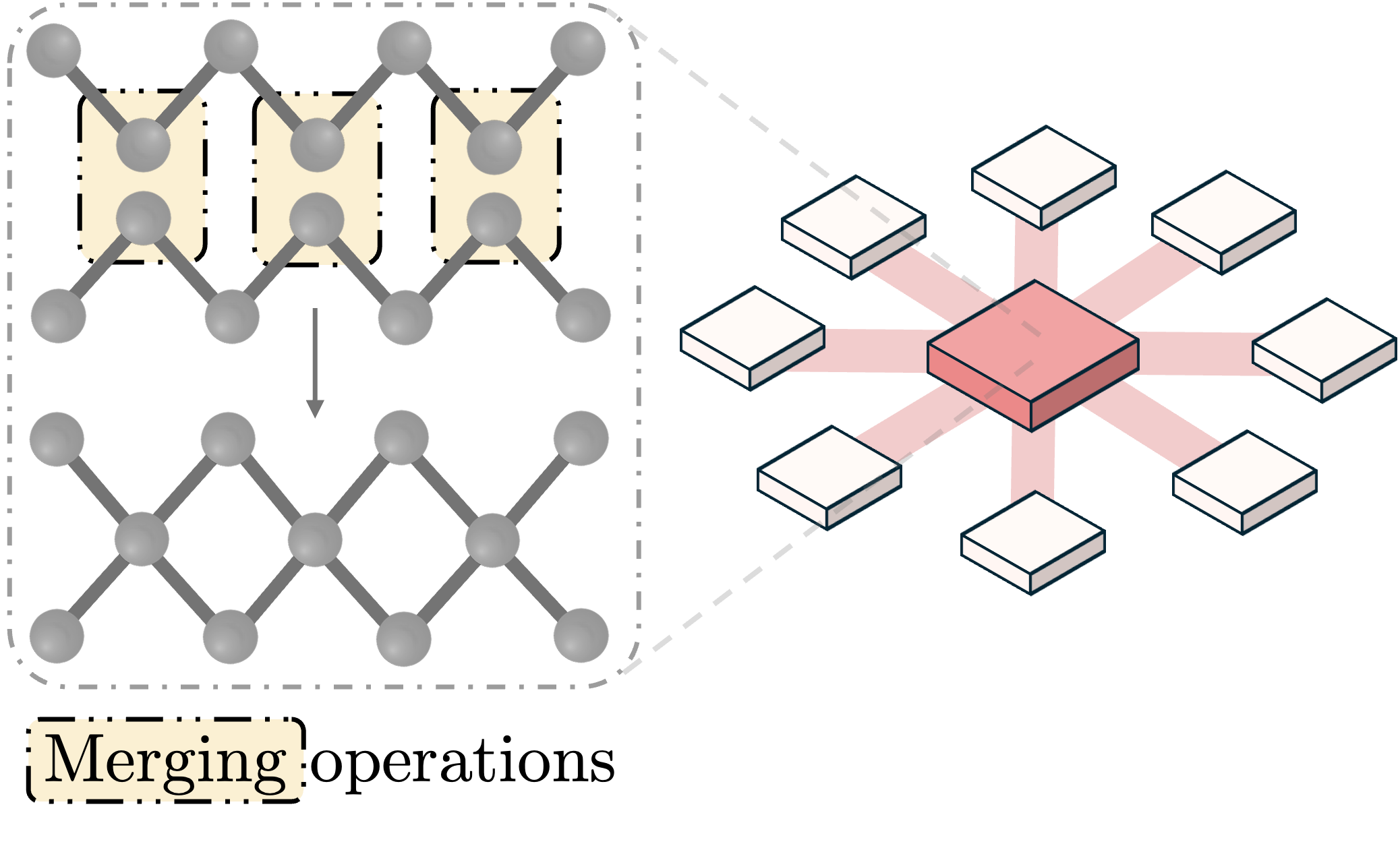}
        \subcaption{Merging of linear graph states performed at the orchestrator.}
        \label{fig:08a}
    \end{subfigure}
    \hfill
    \begin{subfigure}[b]{0.48\textwidth}
        \centering
        \includegraphics[width=1\textwidth]{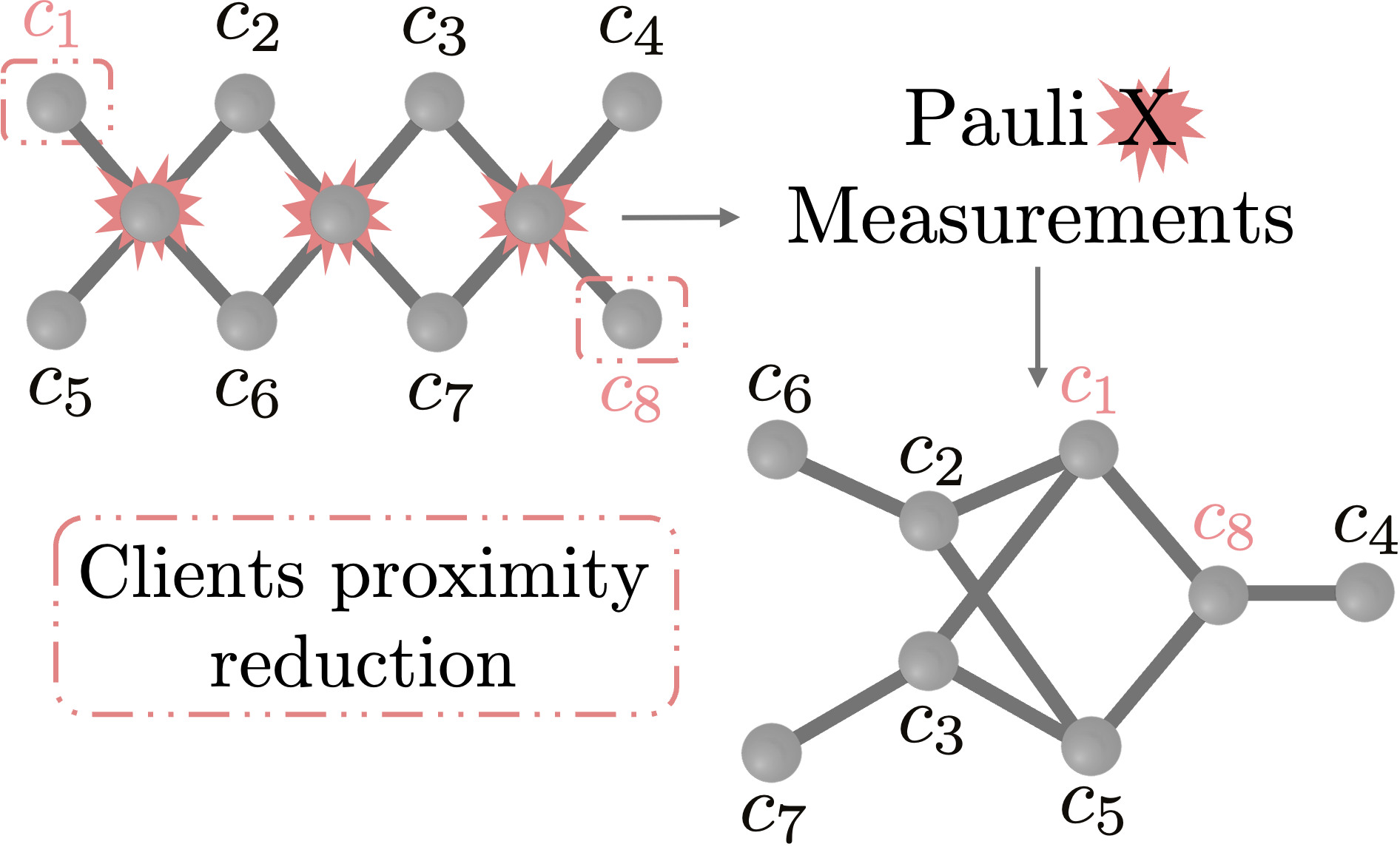}
        \subcaption{Proximity reduction between two clients $c_1$ and $c_8$.}
        \label{fig:08b}
    \end{subfigure}
    \caption{Example of linear graph state manipulations performed at the orchestrator (a) and client proximity reduction through Pauli-X measurements (b).}
    \hrulefill
    \label{fig:08}
\end{figure}

An example of dense resulting artificial topology is represented in Fig.~\ref{fig:06}(b). Remarkably, the presence of multiple entangled links in the artificial topology does not necessarily imply the extraction of more Bell states \cite{MazCalCac-25}, but rather simplifies the selection of the identities of the target nodes for dedicated link extractions. As a consequence, the orchestrator node is capable of enhancing the reliability and adaptability of the overlying entangled network topology, thanks to the possibility of reconfiguring paths according to the traffic demands -- requests of specific entangled links -- or node disconnections. Interestingly, as also represented in Fig.~\ref{fig:08a}, the orchestrator is able to generate and distribute wisely manipulated linear graph states through merging\footnote{The merging operation between vertices belonging to different graph states combines two vertices into a single one. This operation is equivalent to the so-called \textit{fusion} when referred to optical graph state operations.} operations.
Specifically, the orchestrator is able to further engineer the resulting connectivity of the network nodes according to proper Pauli-X measurements. As depicted in Fig.~\ref{fig:08b}, the adjacency between any two target clients can be achieved by reducing the distance of the clients in the resulting artificial topology. As a result, proximity engineering of physically unconnected clients has the direct consequence of simplifying the extraction of the dedicated entangled links in the artificial topology. 

\section{Beyond Quantum Data Centres: Quantum Hubs}
\label{sec:04}

\begin{figure*}[t]
    \centering
    \begin{minipage}[t]{\textwidth}
        \begin{subfigure}[t]{0.38\textwidth}
            \centering            \includegraphics[width=\textwidth]{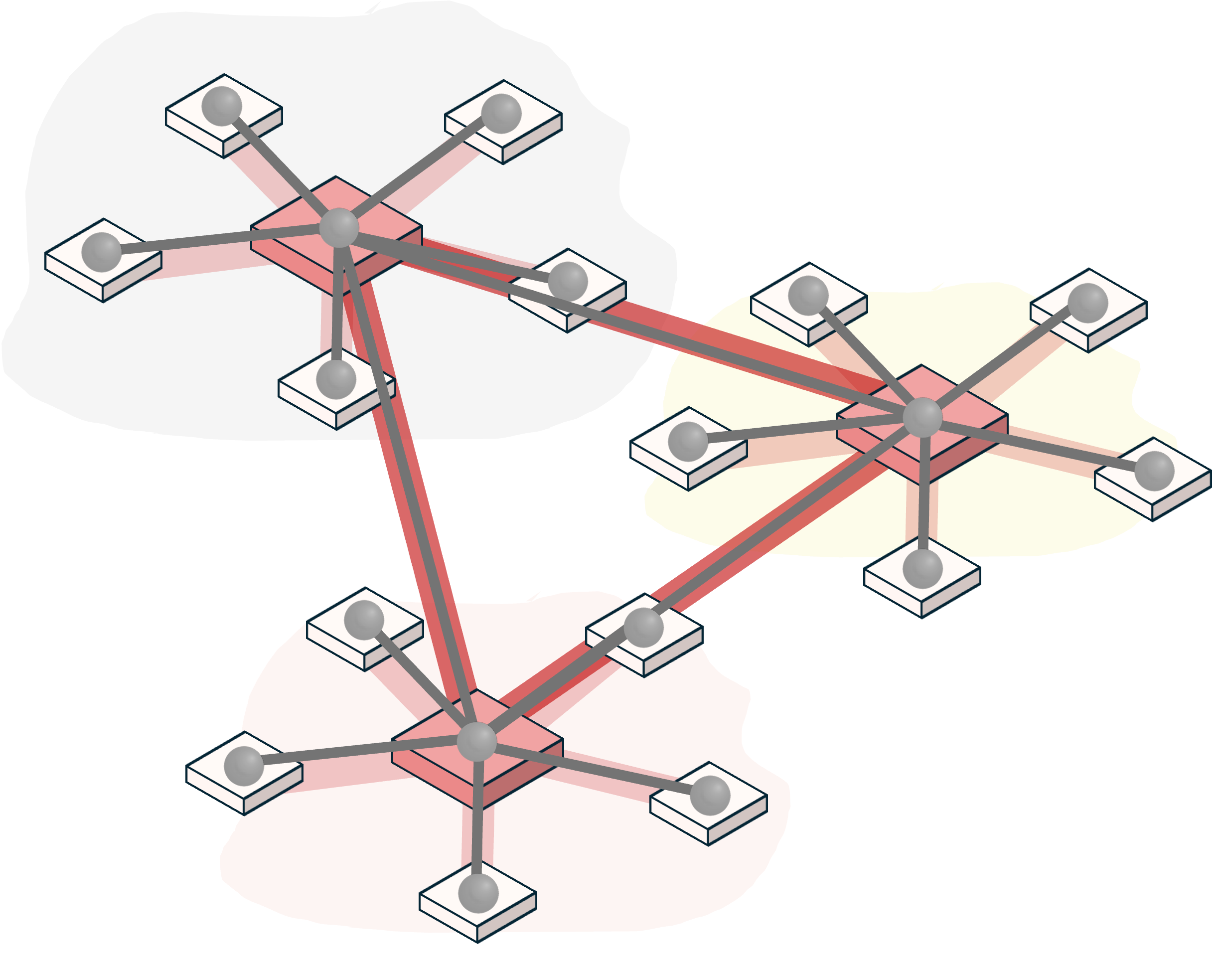}    \subcaption{Schematic representation of the considered physical quantum network architecture in~\cite{CheIllCac-24}. The network comprises several QLANs. Within each QLAN, a orchestrator generates and distributes multipartite entangled states to a set of quantum nodes -- referred to as \textit{clients} -- with a star-like topology. Inter-QLAN connectivity is enabled by point-to-point quantum channels, remarked in dark red, interconnecting different orchestrators.}
            \label{fig:09-a} 
        \end{subfigure}
        \hspace{0.5pt}
        \begin{minipage}[t]{0.3\textwidth}
            \begin{subfigure}[t]{\textwidth}
                 \centering \includegraphics[width=0.85\textwidth]{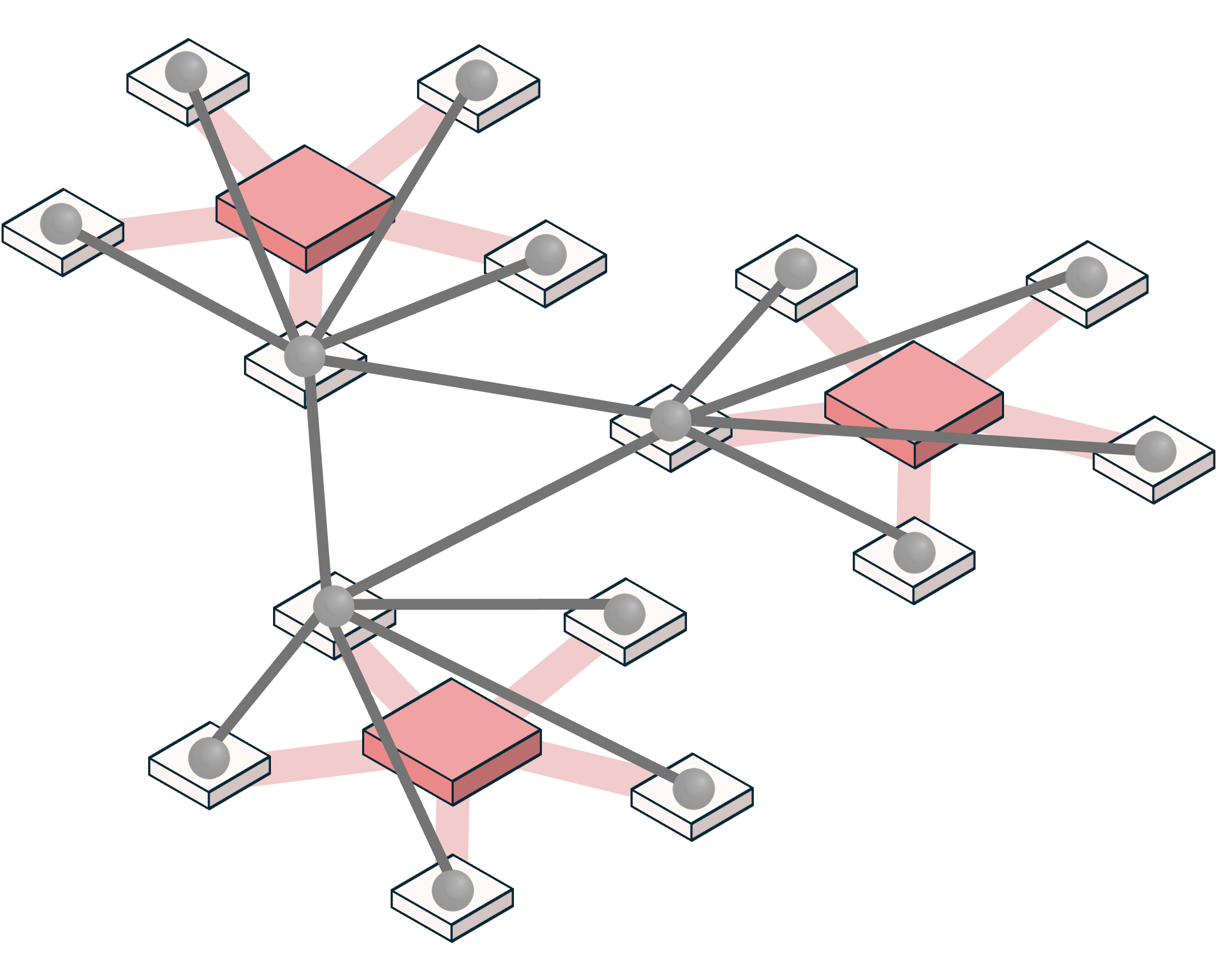}
                 \subcaption{\textit{Role delegation} topology}
                 \label{fig:09-b}%
            \end{subfigure}
            \begin{subfigure}[t]{\textwidth}
                \centering
                \includegraphics[width=0.85\textwidth]{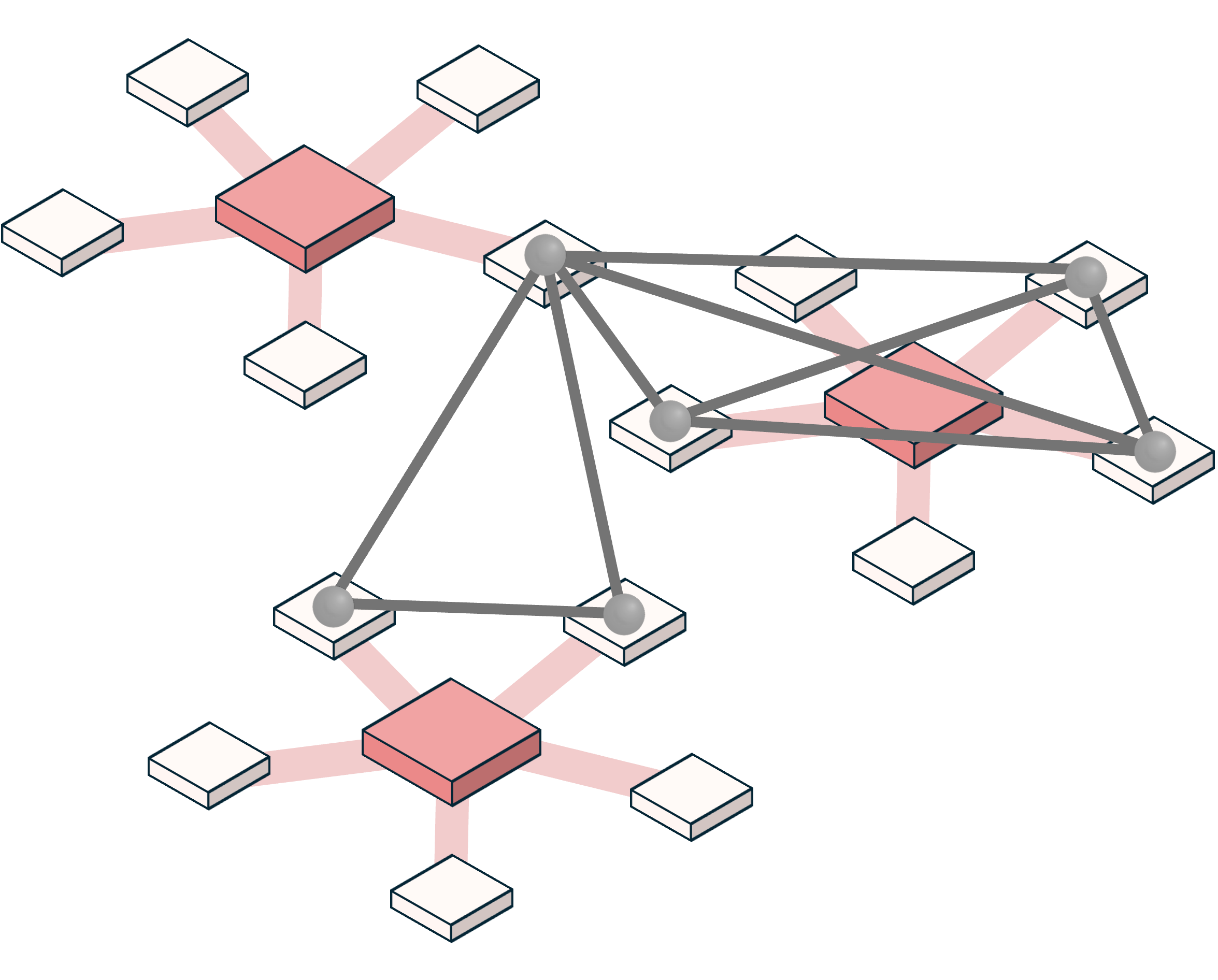}
                \subcaption{\textit{Peer-to-peer} topology}
    	   \label{fig:09-c}
            \end{subfigure}        
        \end{minipage}
        \begin{minipage}[t]{0.3\textwidth}
            \begin{subfigure}[t]{\textwidth}            
               \centering
               \includegraphics[width=0.85\textwidth]{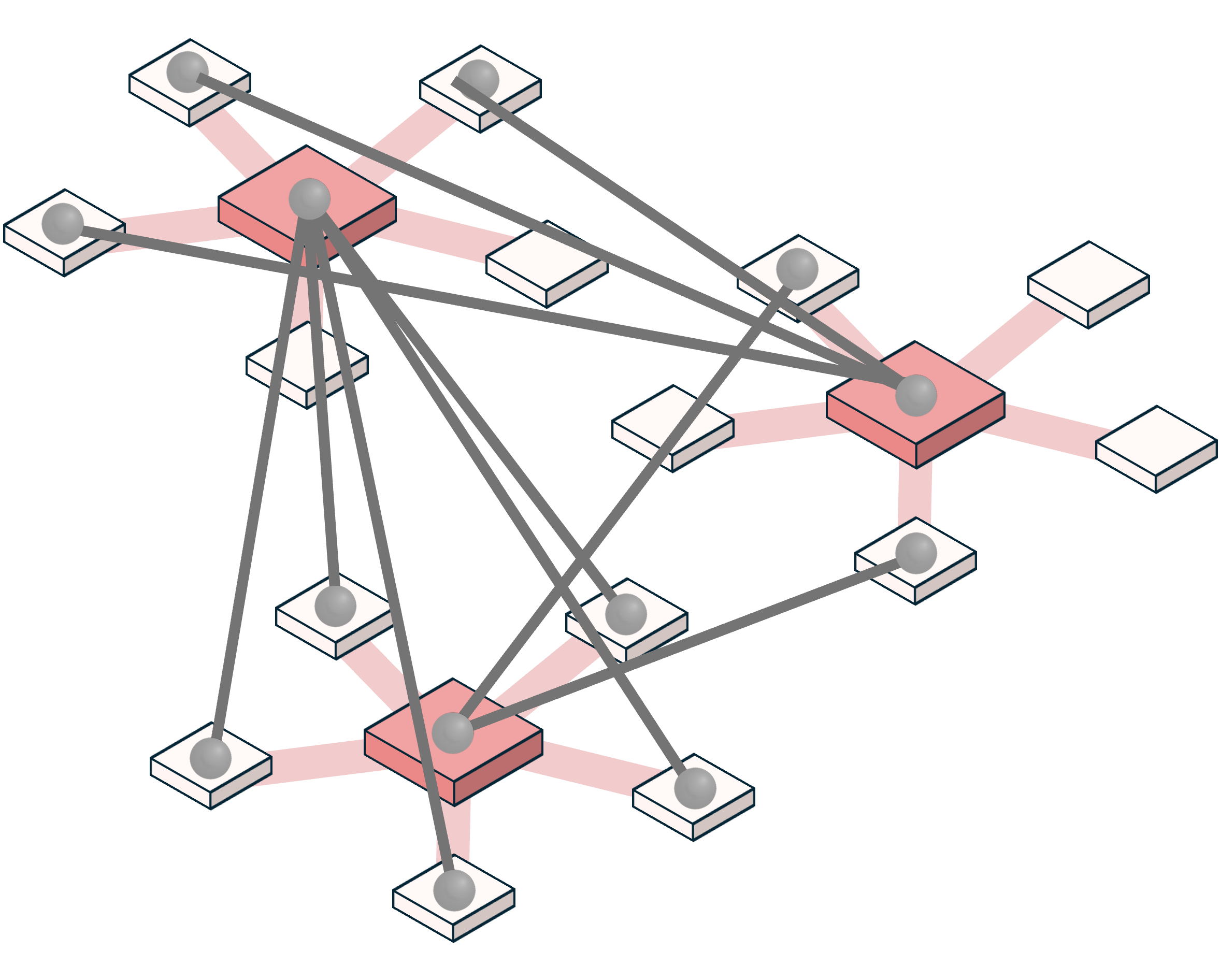}
               \subcaption{\textit{Clients hand-over}  topology}
    	   \label{fig:09-d}
            \end{subfigure}
            \begin{subfigure}[t]{\textwidth}
               \centering
               \includegraphics[width=0.85\textwidth]{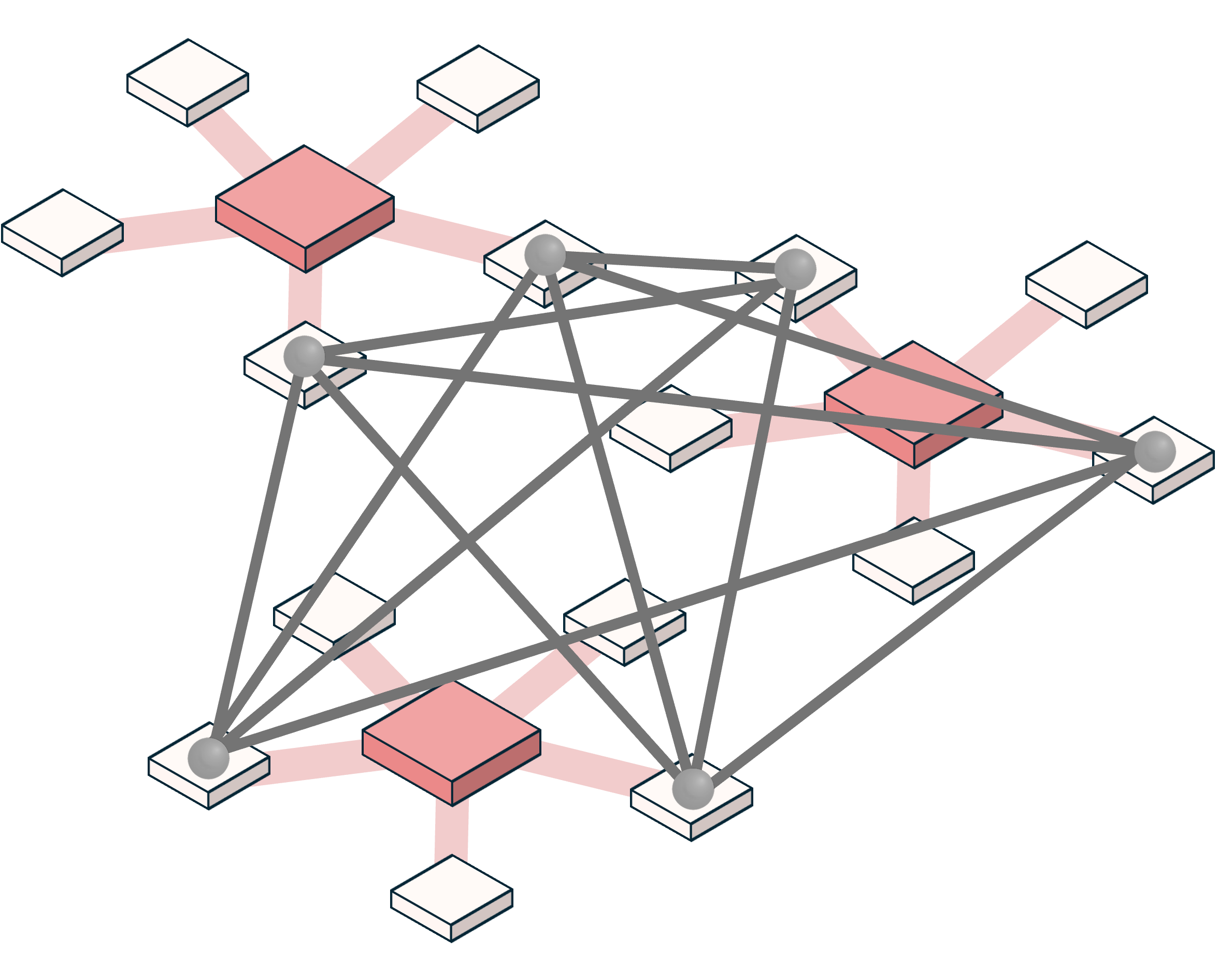}
               \subcaption{\textit{Extranet} topology.}
    	   \label{fig:09-e}
            \end{subfigure}
        \end{minipage}
    \end{minipage}
    \vspace{8pt}
    \caption{Pictorial representation of the inter-QLAN model with three QLANs. Further works to primary archetypes, i.e., \textit{role delegation}, \textit{client hand-over}, \textit{peer-to-peer}, and \textit{extranet} artificial topology, obtained by manipulating a $n$-star graph state with only local operations and measurements.} 
    \label{fig:09}
    \hrulefill
\end{figure*}
In this Section, we provide insights for scaling the interconnection of quantum processors beyond Quantum Data Centres, towards the Quantum Hub archetype. 

As described in Sec.~\ref{sec:02}, to move beyond Quantum Data Centres we need to engineer a network interconnecting different QLANs, by leveraging the connectivity activated by entanglement. 

For the reasons highlighted in Sec.~\ref{sec:03}, each QLAN exhibits a centralized architecture, where the orchestrator handles the entanglement generation and distribution to the QLAN client nodes. This hierarchical design naturally extends to multi-Quantum Data Centre networks, where inter-network physical connectivity is achieved through a mesh topology among the QLAN orchestrators, as represented in Fig.~\ref{fig:09-a}, while the inter-node communications leverage the entanglement-enabled connectivity. Indeed, this approach assures operational consistency across network scales, while respecting the constraints imposed by the technology maturity. As a consequence, from a quantum communication standpoint, the inter-node communication capabilities correspond to the generation of artificial links among network nodes belonging to different QLANS. 
Hence, as for intra-QLAN entanglement connectivity, the choice of the initial multipartite entangled state distributed in each QLAN is of paramount importance, for being able to adapt to different communication patterns \cite{CheIllCac-24-QCE,CheCacChe-23,WanRahRui-24}. 

In this context, \cite{CheIllCac-24} proposes four different prototypes to cater for different traffic patterns and connectivity requirements between QLANs: \textit{peer-to-peer}, \textit{role delegation}, \textit{clients hand-over} and \textit{extranet}. Each of these prototypes is associated with a different artificial topology designed to fulfill specific communication needs between two QLANs. Notably, these topologies are capable of dynamically creating multiple artificial links between distant nodes through local operations only (i.e., using free operations in the sense of quantum communication), without requiring new physical connections. This approach enhances the flexibility and adaptability of quantum networks while maintaining their resource efficiency. We strategically trade off these four traffic pattern prototypes depending on application requirements and network conditions, as follows.

\begin{itemize} 
    \item \textit{Peer-to-Peer}: 
    The ``Peer-to-Peer'' artificial topology is envisioned to be particularly advantageous whenever no information is available on the actual client traffic features. 
    Indeed, the ``hierarchical'' peer-to-peer artificial topology accounts for a hierarchy in terms of hardware requirements between clients and orchestrator. On the other hand, ``pure'' peer-to-peer topology is feasible for distributed network functionalities relying on clients communication capabilities. Specifically, if a client may equally need to communicate with clients belonging to the same QLANs or with a client belonging to a different QLAN, then the communication request will be ready to be served by proactively manipulating the ``pure'' peer-to-peer artificial topology. Indeed, if a client needs to communicate with a client belonging to a different QLAN, then -- by proactively manipulating the artificial topology -- the communication request is ready to be served, without further orchestration at the orchestrator.\\
    
    \item \textit{Role delegation}: Due to the particular structure of the ``role delegation'' artificial topology, a client node, rather than the orchestrator, serves as centre of the star graph. As the name suggests, this topology delegates to a client node the role of orchestrator. Thus, it comes in handy whenever the traffic pattern likely involves a specific node.\\
    
    \item \textit{Clients hand-over}: Different from the above topologies, in the "clients hand-over" topology, artificial links are built between an orchestrator of one QLAN and the clients of a different QLAN. This, from a topological perspective, is equivalent to virtually \textit{move} the clients of a QLAN into a different QLAN, resembling thus a sort of \textit{clients hand-over} from one QLAN to the other. This pattern can be used in quantum network based on trusted relay nodes. If one orchestrator is overloaded or fails, another can take over its clients. \\
    
    \item \textit{Extranet}: ``Extranet'' is fully adapted to inter-domain quantum networks. Specifically, artificial links are created among clients belonging to different QLANs. Thus, inter-QLANs communication needs can be promptly fulfilled, by selecting on-demand -- i.e. accordingly to the current communication request -- the identities of the clients sharing the ultimate artificial link. The degrees of freedom in selecting these identities are higher. This comes without paying the price of additional quantum communications, but only by engineering the proper local operations to be performed at the orchestrator. 
\end{itemize}

These four prototypes can all be realized between two different QLANs, as demonstrated in \cite{CheIllCac-24}. Moving forward, it becomes essential to consider how such a state can be engineered to: i) interconnect nodes belonging to different multiple QLANs, and ii) dynamically adapt to different inter-QLAN traffic demands. For example, how should we address the challenge of increasing the number of QLANs with maintaining the appropriate artificial topologies? How do the four pattern prototypes perform in a multi-QLAN architecture? To this end, we present the following preliminary ideas. By increasing the number of QLANs and nodes, we aim to keep the topology consistent while adjusting the entanglement resources accordingly. One possible direction is to explore the use of more scalable entangled resources — for example, extending from a bi-star state to an $n$-star graph state, as conceptually illustrated in Fig.~\ref{fig:09-b}-~\ref{fig:09-e} for the case of $n=3$. In this extended scenario, it remains an open question whether the artificial inter-QLAN topologies corresponding to primary archetypes — \textit{peer-to-peer}, \textit{client hand-over}, \textit{role delegation}, and \textit{extranet} — can be implemented using only local operations and measurements on such $n$-star entangled states. Investigating this possibility forms an important next step toward scalable and efficient multi-QLAN quantum network architectures.

\section{Open issues and Conclusions}
\label{sec:05}

In this Section, we conclude the paper with some insights on open research directions. Specifically, in addition to the future perspective on the interconnection of Quantum Data Centres presented in Sec.~\ref{sec:04}, some goals that appear common to the whole Distributed Quantum Computing landscape can be recognized.\\

\textbf{Entanglement generation and transduction.} Within the entanglement generation functional block a crucial spot is reserved to quantum transducers. Quantum transducers are essential for the development of heterogeneous networks and for the optimization of quantum computation and communication. Their performance in quantum communication is mainly characterized by the conversion efficiency.
However, it is worthwhile to stress that the transducer efficiency does not provide sufficient granularity to grasp all the mechanisms and phenomena involved in the transduction process, such as the specific type of encoding used within the transducer. Future work is needed to define a standard that takes into account hardware and communication challenges.

\textbf{Entanglement distribution and classical overhead.} 
In the whole distributed quantum computing landscape, classical communication is indispensable for coordinating the distributed quantum operations and measurements. An urgent open question is how classical communication requirements (including latency, synchronization, reliability and coordination overhead) affect the distribution process and to what extent the performance of the virtual quantum processor are affected. Additionally, further investigation is required specifically to investigate the effects of classical overhead for entanglement distribution on the scalability of quantum network topologies. Since quantum operations rely on timely classical signaling to maintain entanglement fidelity, future work should investigate how these classical constraints affect the design and feasibility of dynamic quantum networks.

\textbf{Multi-purpose entanglement utilization.} A central motivation for adopting an artificial topology is the potential to reduce the reliance on the topology arising from physical channels, by creating artificial links through entanglement manipulation via local operations. However, it remains unclear under which conditions artificial links alone are sufficient, and in which cases additional physical links may still be necessary or more effective. This issue becomes particularly challenging when the utilization of entanglement, i.e., artificial links, involves multi-purpose entities, such as archetypes dedicated to running different applications. Future work should rigorously characterize the conditions under which a hybrid strategy (combining physical and artificial links) provides the best trade-off in terms of resource cost, performance and scalability with respect to the entanglement utilization required.

\textbf{Quantum Compiler.} The design of a robust, efficient, and reliable compiler for the distributed quantum computing landscape represents a pressing and unresolved challenge, recognized across multidisciplinary communities. As with many complex system, the choice of the design strategy is shaped by the long-term vision for the evolution of the distributed quantum computing landscape. 
In this regard, a particularly critical aspect concerns the interplay between the compiler and the network entities, with respect to the entanglement utilization, generation and distribution. 
At one end of the design spectrum, the compiler -- stemming from the algorithm to be executed -- is given full control over the process of entanglement generation and distribution. Hence, in this algorithm-driven approach the compiler instructs the quantum network according to its computational needs. At the opposite end, the network is responsible for managing entanglement independently, and the compiler must adapt to the available network resources. Within this network-driven approach, a range of possibilities emerges: from a static model where the network topology is provided as a fixed input to the compiler, to more dynamic scenarios where some form of negotiation or real-time interaction between the compiler and the network is established. This flexibility introduces a key design point that calls for a careful co-design between network protocol and compiler strategies in order to ensure efficient and scalable execution of distributed quantum algorithm. Clearly, the algorithm-driven approach appears as a simple and natural choice for Multi-Core archetype, as the low-complexity scales allow for an ad-hoc management of the entangled resources. However, it is still unclear how this co-design should be developed in the Quantum Data Centre archetype and beyond. \\

In conclusion, in this work we discussed how quantum networks enable the distributed quantum computing landscape at different scales of complexity and heterogeneity. The role of entanglement in such networks has been highlighted both for sending qubits (TeleData) and executing operations on them (TeleGate). Different architectural models for interconnecting quantum computers have been presented, with particular emphasis on the Quantum Data Centre, a network that interconnects small number of quantum computers, which is expected to represent the key technological building block in the midterm. We deeply discussed and recognized that one of the major challenges in interfacing currently available quantum technologies lies in the problem of quantum transduction, i.e., the process of conversion between superconducting qubits into flying qubits. Moreover, we provided a communication engineering perspective on Quantum Data Centres by discussing the structural properties and topological constraints of QLANs. This highlighted how the use of an orchestrator responsible for generating and distributing entanglement among nodes, allows to dynamically create different network topologies via local measurements, by unlocking possibilities beyond the physical interconnection of the computing units. In addition, future developments have been presented, with a focus on the communication perspectives of interconnecting multiple Quantum Data Centres towards the development of Quantum Hubs. Finally, open issues that must be still addressed in order to build a fully functional Quantum Data Centre have been discussed.

\section*{Acknowledgement}{This work has been funded by the European Union under the ERC grant QNattyNet, n.101169850. Views and opinions expressed are however those of the author(s) only and do not necessarily reflect those of the European Union or the European Research Council. Neither the European Union nor the granting authority can be held responsible for them.}


\bibliographystyle{ieeetr}
\bibliography{Bibliography} 

\end{document}